\documentclass[%
 reprint,
 amsmath,amssymb,
 aps,
]{revtex4-1}

\usepackage{color}
\usepackage{graphicx}% Include figure files
\usepackage{dcolumn}% Align table columns on decimal point
\usepackage{bm}% bold math
\usepackage[pdftex,plainpages=false,colorlinks=true,linkcolor=blue, citecolor=blue, urlcolor=blue]{hyperref}% add hypertext capabilities
\begin{document}

\preprint{APS/123-QED}

\title{Superconductivity in the Two-Orbital Hubbard Model of Infinite-Layer Nickelates}

\author{Zhihui Luo}
\affiliation{Center for Neutron Science and Technology, Guangdong Provincial Key Laboratory of Magnetoelectric Physics and Devices, School of Physics, Sun Yat-sen University, Guangzhou, Guangdong Province 510275, China
}

\author{Dao-Xin Yao}
\email[Corresponding author: ]{yaodaox@mail.sysu.edu.cn}
\affiliation{Center for Neutron Science and Technology, Guangdong Provincial Key Laboratory of Magnetoelectric Physics and Devices, School of Physics, Sun Yat-sen University, Guangzhou, Guangdong Province 510275, China
}

\author{W\'ei W\'u}
\email[Corresponding author: ]{wuwei69@mail.sysu.edu.cn}
\affiliation{Center for Neutron Science and Technology, Guangdong Provincial Key Laboratory of Magnetoelectric Physics and Devices, School of Physics, Sun Yat-sen University, Guangzhou, Guangdong Province 510275, China
}

\date{\today}% It is always \today, today,

%\tableofcontents

\begin{abstract}
The pairing symmetry in infinite-layer nickelate superconductors has been an intriguing  problem under heated debates. In this work,
we study a two-orbital Hubbard model with one strongly correlated $3d$ orbital and one more
itinerant $5d$ orbital, by using an eight-site cellular dynamic mean field
theory study. We  establish a superconducting phase diagram with  $d_{x^{2}-y^{2}}$,
$s_{\pm}$ and $d+is$ wave pairing symmetries, based on which we clarify the roles of various relevant parameters including hybridization
$V$, itinerant carrier density $\langle n_{c}\rangle$ and interaction
$U_{c}$. We show that the inclusion of a less correlated $5d$ band in general suppresses the $d_{x^{2}-y^{2}}$ wave pairing.
We demonstrate that  the $d+is$ wave is maximized when the $5d$ orbital has a large Coulomb repulsion with intermediate hybridization parameter.
We perform fluctuation diagnostics to show that the driving force behind the $d_{x^{2}-y^{2}}$ wave is the intraband antiferromagnetic fluctuations in the $3d$ orbital, while for the $s_{\pm}$ wave, the pairing is mainly from the antiferromagnetic correlations residing on the  local $3d$-$5d$ bond  in real space.
\end{abstract}


\maketitle

\section{\label{sec:1}Introduction}

Identifying the predominant pairing symmetry is a fundamental problem in  the study of unconventional
superconductors which is usually rooted in comprehending the intricate underlying
mechanism of electron pairings.  In the case of cuprate superconductors, the spin-singlet pairing with $d_{x^{2}-y^{2}}$ wave
symmetry dominates the superconducting (SC) phase diagram,
which is linked to doping a Mott insulator 
%with strong antiferrogmagnetic (AFM) correlations
on an effective  single band two-dimensional square lattice~\citep{patrick06}.
In iron-based superconductors (FeSC),  $s_{\pm}$  wave pairing
is ubiquitous in various compounds, due to the multi-orbital physics of  the intricate Fe-3$d$ orbitals at the
Fermi energy ($E_{\rm F}$) that favours the inter-band $s_{\pm}$ wave pairing. It is notable that in some pnictides or chalcogenides, a proximate $d_{x^2-y^2}$ wave pairing may also be possible
\citep{leewc09,Fernandes13} when adjusting the pnictogen height \citep{platt12}
or  doping \citep{maier11_d}.
%, which is achieved by varying  Fermi surface.
%It's worth mentioning that, both cuprates and FeSC show inextricably bound to antiferromagnetic fluctuation \citep{Scalapino_RMP12}.

Experimentlists recently discovered a new class of  nickelate-based
 superconductors $\mathcal{R}$NiO\textsubscript{2}
($\mathcal{R}$ represents rare-earth element La, Pr or Nd) \citep{Lidanfeng_nature19,Lidanfeng_PRL20,ZengShengwei_PRL20,guqq20,oasda20,osada20_prm,zhouxr21,pangrace21,osada21,shengwei22,zengsw22}.
Despite hosting a relative low SC transition temperature ($T_{c}$) ( $\sim10$ K)
, the nickelate-based superconductors are  widely believed to have
 profound theoretical implications for the field of unconventional superconductivity, given its analog to cuprate superconductors and its strong-coupling nature. Concerning the pairing symmetry,
both nodal and nodeless,
or even nodal+nodeless gaps \cite{guqq20,chow22,harvey22} have
been reported in tunneling or London penetration measurements, implying a possible coexistence of $d_{x^{2}-y^{2}}$ and
$s_{\pm}$ wave pairings. 
%Such complexity is associated to the relative position of the gap node and Fermi surface profile.  
Subsequently,  a theoretical
conjecture was proposed \cite{zfc20} which suggests the existence of a crucial Kondo exchange  in nickelates
that drives the pairing symmetry towards the nodeless $s_{\pm}$ wave.
This proposal points to a multi-orbitals nature of the low-energy  physics of nickelates.
Meanwhile, there is also evidence indicating that  nickelates  have a rather large
charge-transfer energy within the Zaanen-Sawatzky-Allen (ZSA) scheme
\citep{zaanen85,jiangm20,emily20,hepting20,botana20}, which means that,  to some extend, nickelates
may behave as a single-band doped Mott insulator, much like  the cuprates. Nevertheless, 
the undoped nickelates in fact exhibit weak insulating characteristics \citep{Lidanfeng_PRL20,ZengShengwei_PRL20,oasda20,osada21}
which may be explained as the self-doping effect via the Kondo couplings \citep{zfc201}
between the $\mathcal{R}$-$5d$ states and  the 2D hole band of  Ni-$3d_{x^{2}-y^{2}}$
orbital \citep{hepting20,goodge21}.  On the other hand, density functional theory plus dynamical mean-field theory (DFT+DMFT) calculations also emphasize a crucial multi-orbital description of the nickelates \citep{petocchi20,leonov20,karp20,lechermann20_prx,lechermann20,lechermann21}.

The intriguing  phenomena of nickelates have attracted much attention to a very basic question:  how the pairing symmetry is determined in a two-dimensional interacting electron system that has both strongly and less correlated bands (such as the Ni-$3d_{x^2-y^2}$ and $\mathcal{R}$-$5d$ orbitals in nickelates respectively)? Specifically, as the hybridization strength $V$ between the two bands changes, the geometry of Fermi surface (FS) would change accordingly, leading to the immediate alteration of the free energy of the pairing state with certain symmetry. A subsequent question then prompts that whether the geometry of FS is the only important factor in determining the symmetry of pairing \citep{hujp12}? To address these questions, we study a doped two-band Hubbard model to establish a phase diagram describing the  evolution trajectory of the pairing symmetries. We also explore the dependence of the superconducting $T_c$ of $d_{x^2-y^2}$ and $s_{\pm}$ wave pairing on various parameters, such as the strength of the Hubbard $U$ at both bands, the filling factor $\langle n \rangle$ , etc.  Moreover, we search in the parameter space for  the possible 
coexistence of the $d_{x^2-y^2}$ and $s_{\pm}$ wave  instability, namely, the $d+is$ wave. We find that it only exists when the two correlated bands both have a relatively large $U$ and in the intermediate $V$ regime. Finally, we perform
fluctuation diagnostics analysis in the superconducting phases, which reveals that the $d_{x^2-y^2}$ and $s_{\pm}$ wave pairing symmetries
have different origin of pairing:  the $d_{x^2-y^2}$ wave is largely driven by the $(\pi, \pi)$ antiferromagnetic fluctuations whereas for $s_{\pm}$ pairing, is more from local in real-space inter-band antiferromagnetic fluctuations.

\section{\label{sec:2} Model and Methods}

We consider the two-orbital Hubbard model that reads \citep{bulut92,hetzel94,scalettar94,dos95,bouadim08,kancharla07,maier11,mishra16,nakata17,maier19,karakuzu21,nica20}

\begin{eqnarray}
\label{Eq:hamil}
\mathcal{H} & = & \sum_{\rm{k}\sigma}\left(t_{\rm{k}}^{d}-\mu^{d}\right)d_{\rm{k}\sigma}^{\dagger}d_{\rm{k}\sigma}+\sum_{\rm{k}\sigma}\left(t_{\rm{k}}^{c}-\text{\ensuremath{\mu^{c}}}\right)c_{k\sigma}^{\dagger}c_{k\sigma}\nonumber \\
 &  & +V\sum_{\rm{k}\sigma}\left(d_{\rm{k}\sigma}^{\dagger}c_{\rm{k}\sigma}+h.c.\right)\\
 &  & +U_{d}\sum_{i}n_{i\uparrow}^{d}n_{i\downarrow}^{d}+U_{c}\sum_{i}n_{i\uparrow}^{c}n_{i\downarrow}^{c}.\nonumber
\end{eqnarray}

where $d_{\rm{k}\sigma}^{\dagger}(d_{\rm{k}\sigma})$ and $c_{\rm{k}\sigma}^{\dagger}(c_{\rm{k}\sigma})$
are  the creation (annihilation) operators for the more correlated orbital ( $d-$ orbital  ) and  the less correlated $c$-orbitals, respectively.
$V$ is hybridization between $d$ and $c$-orbitals, $U_{d}$ and
$U_{c}$ are the on-site Coulomb repulsion for $d$ and $c$-orbitals.
Without loss of generality, here we consider equal  intra-layer nearest-neighbor
hopping for  $d$ and $c$-orbitals with $t_{\rm{k}}^{d}=t_{\rm{k}}^{c}=-2\left(\cos \mathrm{k}_{x}+\cos \mathrm{k}_{y}\right)$,
which closely resemble the estimated value $t^{\mathrm{Nd}}/t^{\mathrm{Ni}}\approx1.03$ in density
functional theory (DFT) calculation for nickelates \citep{wuxx20}. $\mu^{d}$
and $\mu^{c}$ are the corresponding chemical potentials. Unless specified,
we use the doping for $d$-orbital $p=1-\langle n_{d}\rangle=0.05$
and $U_{d}=8$ throughout the paper.

To solve the above Hamiltonian, we use an eight-site  cellular dynamical mean-field theory (CDMFT), which is a cluster extension
of the dynamic mean-filed theroy (DMFT).
The CDMFT method captures non-local correlations within the cluster exactly,
while longer-range spatial correlations are dealt by a dynamical mean-field
that embedded in an effective cluster impurity model~\citep{maier05,maier05_prl}.
It is worth nothing that in CDMFT  we do not make any presumption of the leading instability like in weak coupling
approaches~\citep{fernandes17} or rescaling of electronic band structure
in exchange model, which is crucial for mapping out the unconventional SC \citep{georges96,kotliar01,maier05}.
To solve the CDMFT impurity problem, we use the Hirsch-Fye quantum Monte Carlo
solver \citep{hirschfye86} with a discrete time $\Delta\tau=0.1$.

\section{\label{sec:4}Result}

\subsection{Evolution from $d_{x^2-y^2}$ to $s_{\pm}$ wave pairing}

\begin{figure}[h]
\includegraphics[scale=0.63]{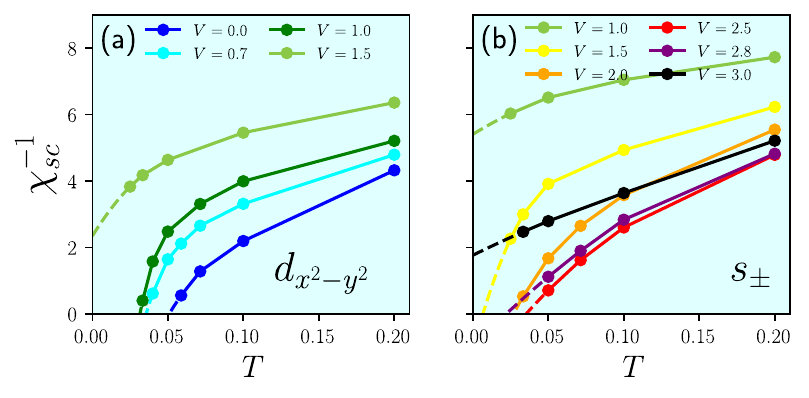}

\caption{\label{fig:2}The inverse pairing susceptibility $\chi_{sc}^{-1}$
as a function of $T$ with varying $V$ for (a) $d_{x^{2}-y^{2}}$
and (b) $s_{\pm}$ wave pairings. The extending dashed curves at low $T$
are quadratic interpolations. }
\end{figure}

We investigate the pairing susceptibility~\citep{lin12,ww14,wuwei15}
$\chi_{sc} \equiv \chi_{sc}^{dd,dd} =1/N\sum_{i,j} \int_0^{\beta} \langle T_{\tau} P_j^{\dagger} (\tau) P_i(0)  \rangle d \tau$, where $P_i = (1/\sqrt{2})f(r)(d_{i, \uparrow}d_{i+r, \downarrow} -d_{i ,\downarrow}d_{i+r, \uparrow})  $  is the spin singlet pairing operator in the $d$-orbital , and $f(r)$ represents the $d$ or $s_\pm$-wave  form factor associated with a pairing bond $r$. In practice, here $\chi_{sc}$ is obtained by computing the linear pairing response of the system with  a small pinning pairing field induced in Eq.~(\ref{Eq:hamil}). We first display the inverse pairing  susceptibility  $\chi^{-1}_{sc}$ as a function of temperature $T$ for a series of different hybridization $V$  in Fig.~\ref{fig:2}(a) ($d_{x^2-y^2}$ wave) and Fig.~\ref{fig:2}(b) ($s_{\pm}$ wave). Here we use a typical set of parameters $U_d = 8, U_c =4, n_d = 0.95, n_c = 0.9$. As one can see that at small $V$, $\chi^{-1}_{sc}$ of  $d_{x^2-y^2}$ wave pairing decreases rapidly with lowering temperature $T$, which extrapolates to zero as approaching a finite temperature that we identify as the superconducting $T_c$. From Fig.~\ref{fig:2}, we see that as $V$ increases, $T_c$ of $d_{x^2-y^2}$ wave pairing decreases monotonously~\citep{jiang20}. In contrast, for $s_{\pm}$ wave the pairing instability first grows with $V$ as $V \lesssim 2.5$, and  $T_c$ reaches its maximum at $V\sim2.5$, then it quickly falls to zero as $V$ approaching $V \sim 2.9$, see Fig.~\ref{fig:2}(b).

\begin{figure}
\includegraphics[scale=0.63, trim=12 0 0 0]{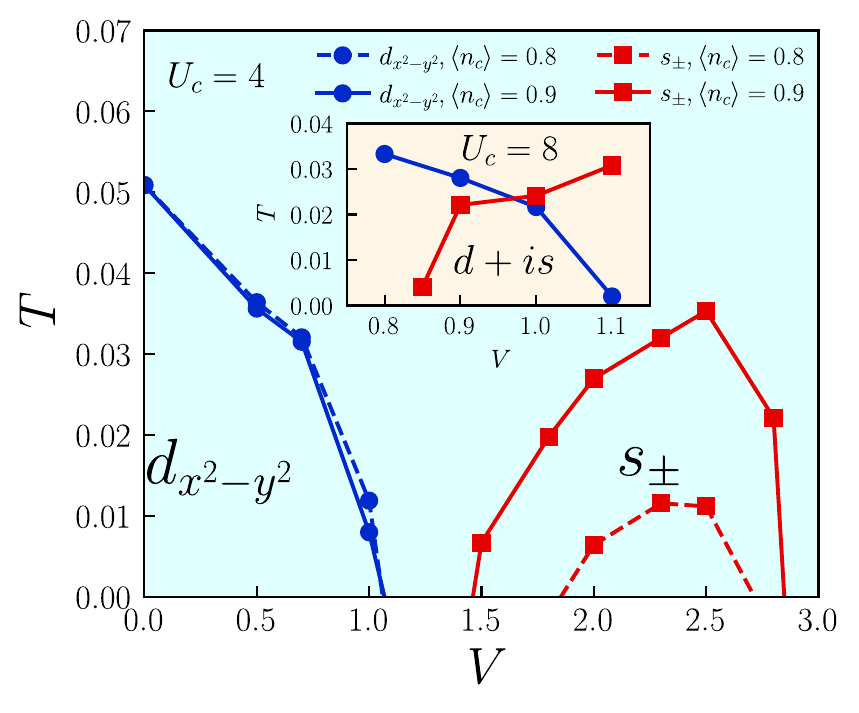}

\caption{\label{fig:1}  $T_{c}$ as a function of $V$ for $n_{c}=0.8$ (dashed
lines) and 0.9 (solid lines) with $U_{c}=4$ (cyan background). The
$d_{x^2-y^2}$ and $s^{\pm}$ wave pairings are marked by blue rounds and red squares.
Inset: $T_{c}$ as a function of $V$ for $U_{c}=8$ (light orange
background), with $\langle n_{c}\rangle=0.9$.}
\end{figure}

 Repeating above computations for different $\langle n_{c}\rangle$,
$U_{c}$ parameters, we obtain phase diagrams in Fig.~\ref{fig:1}(a), which display $d_{x^{2}-y^{2}}$ and $ s_{\pm}$ superconducting $T_{c}$ as functions of $V$. The solid curves in the main plot
show result for $U_{c}=4$ (and $U_d = 8$) ,$\langle n_{c}\rangle=0.8$,
 which reveals a clear evolution of the dominant pairing symmetry from $d_{x^{2}-y^{2}}$
to $s_{\pm}$ wave as increasing $V$. Namely,  for small $V$ ($V\lesssim1.0$),
the $d_{x^{2}-y^{2}}$ pairing (dots) prevails,
while for stronger hybridization $1.5\lesssim V\lesssim2.8$, the system becomes prone to
  $s_{\pm}$ wave pairing (squares). Finally, when $V$ is further increased, $i.e.$, when $V>2.8$, the  $s_{\pm}$ wave $T_{c}$  rapidly drops to zero, as localized singlets formed between $d$ and $c$-electrons in a unit cell~\cite{maier11}.

%\out{ Generally speaking, for $d_{x^{2}-y^{2}}$ pairing, increasing $V$ overall suppresses $\chi_{sc}$, but for $s_{\pm}$ pairing, $\chi_{sc}$ is also enhanced and develops finite $T_{c}$ at $V\sim1.5$ (yellow), then reaches its maximum at $V\sim2.5$ (red), and again diminishes with further increasing $V$. We note that from $V=2.5$ to $3.0$, the suppression of $\chi_{sc}$ follows with a flattening trend of $\chi_{sc}^{-1}(T)$ curves, which indicates the change of the scaling  from  exponential to power-law growth,   and further suggests a total disfavor of $s_{\pm}$ pairing for large $V$ neither supported by single-particle excitation nor two-particle fluctuation.}

%\ww{A sentence describing the importance of the value of $\langle n_{c}\rangle$ in Ni-base superconductors }
In addition to $V$, the pairing instabilities can 
 also be sensitivity to the $c$-orbital density $\langle n_c\rangle$.
To  investigate the doping effects of $c$-orbital on SC, we compare the dashed lines ($\langle n_c \rangle =0.8$) with  the solid lines ($\langle n_{c}\rangle=0.9$) in the main plot of  Fig.~\ref{fig:1}, where  one can see that changing $c$-density  $\langle n_{c}\rangle$ does not significantly affect $d_{x^{2}-y^{2}}$ wave pairing, whereas the
$s_{\pm}$ pairing does demonstrate great variation  to  $\langle n_c \rangle$~\citep{nica20,karakuzu21}.
For example, at $V=2.5$, $T_{c}$  of $s_{\pm}$ wave pairing exhibits a  drastic drop from 0.035
to 0.008 as $\langle n_{c}\rangle$ decreases  from $\langle n_{c}\rangle=0.9$ to $\langle n_{c}\rangle=0.8$.
In our study, when further decreasing  $\langle n_{c}\rangle$, we could not find a finite $T_c$ for $s_{\pm}$ wave pairing
in the temperature range that accessible to our study.
To understand this remarkable difference in the dependence on $\langle n_{c}\rangle$ between $d_{x^{2}-y^{2}}$ and $s_{\pm}$ wave pairing,
we first note that for the $d_{x^{2}-y^{2}}$ wave pairing, it is typically associated with correlations between electrons effectively in a single band system~\cite{zrs}.  Here we use  $U_d = 8, U_c=4$ such  that  the correlations effects in $c$-orbital is too weak to drive a $d_{x^{2}-y^{2}}$ wave instability (In CDMFT on single band Hubbard model, $U>6$ is usually required to obtain superconductivity~\cite{fratino16}). Thus the changing of $n_c$ should not change much the $d_{x^{2}-y^{2}}$ wave pairing that  caused by the $d$-electrons at small $V$. For
the  $s_{\pm}$ wave pairing, since both $d$ and $c$-orbitals are involved, the situation can be more complicated. Below we will tackle this problem from two aspects: (1)  We study the changing of Fermi surface with $V$ and $n$, which represents a typical low-energy property of the system that immediately affects the free energy of the  pairing. (2) We employ the fluctuation diagnostics technique \citep{Gunnarsson15} to investigate the  ``pairing glue" between electrons that usually associated with higher energy aspect of the system.

%\out{ Such a diversity could be intuitively explained as single-layer preferred $d_{x^{2}-y^{2}}$ pairing versus bilayer preferred $s_{\pm}$ pairing. For the $d$ pairing dome, $T_{c}$ has already notably suppressed by inter-layer $d-c$ hybridization $V$ before it can sufficiently senses the influence from $\langle n_{c}\rangle$; For the $s_{\pm}$ pairing dome, however, $\langle n_{c}\rangle$ is crucial for providing effective hybridization, thus even a slight decrease of $\langle n_{c}\rangle$ can results in a suppression of pairing. }

\begin{figure}
\includegraphics[scale=0.52,trim=15 0 0 0]{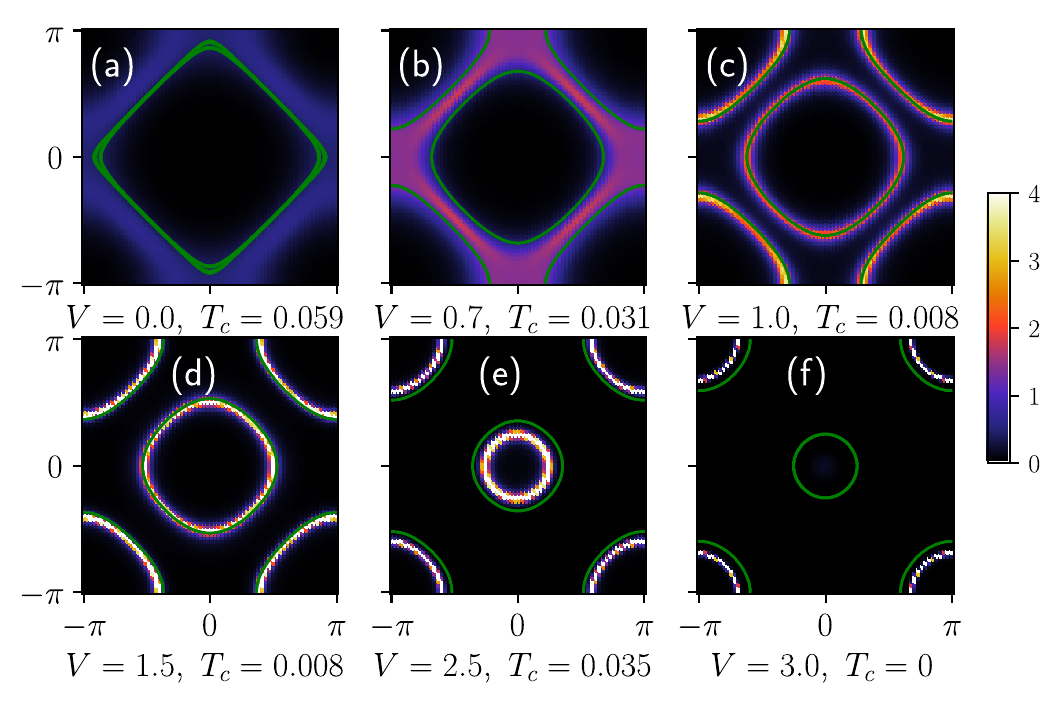}

\caption{\label{fig:3} Interacting Fermi surface in momentum space.  (a-c) For small $V$  in the $d_{x^2-y^2}$ pairing regime, with $V=0.5,\ 0.7$ and 1.0.
Note that the $d_{x^{2}-y^{2}}$ wave $T_c$ at these $V$ is labeled in the respective plots.
(d-f) For lager $V$  in the $s_{\pm}$ wave pairing regime, with $V=1.5,\ 2.5$ and $3.0$. Here the dominant  $s_{\pm}$ wave  $T_c$ at these $V$ are labeled in the  respective plots. For reference, the green lines indicate the corresponding non-interacting FS. Here $\langle n_{c}\rangle=0.9$, $U_{c}=4$, $T=1/30$. }
\end{figure}

%\out{ The above features suggest that, increasing $V$ follows with an evolution of spin fluctuation from intra-layer to inter-layer \citep{maier11,nica20,karakuzu21}, eventually leads to an evolution from $d_{x^{2}-y^{2}}$ to $s_{\pm}$ pairing. From the perspective of single-particle excitation, $s_{\pm}$ pairing is associated with pair scattering between electron and hole pockets, formed respectively by the bonding and anti-bonding of the $d$ and $c$ electrons. }

 Fig.~\ref{fig:3} presents the interacting Fermi surface along with non-interacting FS  shown in green
lines.  The upper panel in Fig.~\ref{fig:3}(a-c) is for smaller $V$ cases,  $V=0.5,0.7,1.0$ where $d_{x^{2}-y^{2}}$
wave can be hosted at low $T$ in this regime, and the lower panel in Fig.~\ref{fig:3}(d-f) is for larger $V$ cases,  $V=1.5,2.5,3.0$ where $s_{\pm}$
pairing occurs at low $T$ in this regime.  As one can see that, as increasing $V$,
the FS as a whole evolves from diamond-like shape to  a  well separated electron-pocket/ hole-pocket structure.
This is generally in accordance with the evolving of the pairing symmetry from  $d_{x^{2}-y^{2}}$ to $s_{\pm}$ wave, as the former prefers
a diamond-like Fermi surface to minimize the free energy, the latter is more  associated with pairing electrons scattered between electron and hole pockets~\cite{hjp_SR2012}.
There are a couple of points we would like to stress on: (1)
Putting aside the blur (continuum) between $d-c$ bands in momentum space at small $V$, generally speaking, interaction does not change much the shape or loci of the Fermi surface at large $V$. In the zero $V$ (or the single-band) limit, the shape of the Fermi surface can be strongly renormalized at small dopings, due to the particle-hole asymmetry in the electron scattering~\citep{wuwei18}. Here we see that for $V \geq 0.7$, in contrast, the interacting Fermi surface is in generally outlined by its non-interacting counterpart. (2) In the intermediate $V$ regime, \textit{i.e.}, $V \sim 1.0$ in Fig.~\ref{fig:3}(c), the geometry of the Fermi surface in principle allows both  $d_{x^{2}-y^{2}}$ and  $s_{\pm}$ wave pairing with minimized free energies. In other words, for both  gap functions $\Delta(\rm{k})$,  there is  a large overlap between the non-zero $\Delta(\rm{k})$ and the Fermi surface \cite{hjp_SR2012}, which  makes the pairing energetically possible.  Moreover, here  scattering of electrons within the same band or between the two different bands are both  feasible, which does not prevent the the form of  $d_{x^{2}-y^{2}}$ or $s_{\pm}$ wave  Copper pairs. However, in our study we  in fact did not  detect the sign of the coexistence of  $d_{x^{2}-y^{2}}$ and  $s_{\pm}$  pairing in this parameter regime. Namely, here we only found a finite $T_c$ for  $d_{x^{2}-y^{2}}$ wave pairing, but no finite $T_c$ for  $s_{\pm}$ pairing is seen in the  temperature range accessible to us. As we will show below,  the value of $U_c$ is  critical to the formation of  $s_{\pm}$  pairing at intermediate $V$ ($V \sim 1$). We found that only when $U_c \gtrsim  6 $, a finite $T_c$ for  $s_{\pm}$  pairing can be reached  at $V \sim 1.0$. 
This suggests that, besides the  Fermi surface geometry, other ingredients like the subtle change of Hubbard $U$ between electrons are also crucial in driving the $s_{\pm}$  pairing instability.

Above we have shown results for $\langle n_{c}\rangle=0.9$ in Fig.~\ref{fig:3}.
One may expect that the strong $\langle n_{c}\rangle$ dependence of $s_{\pm}$ wave pairing shown in Fig.\ref{fig:1}
could also manifest in the $\langle n_{c}\rangle$ dependence of FS topology, since decreasing $\langle n_{c}\rangle$
results in the shrink of electron pocket and the expansion of hole pocket
in a way disfavoring $s_{\pm}$ pairing. Here we have verified that such FS deformation is in fact marginal (no shown). We argue that the detrimetal effect on $s_{\pm}$ pairing by decreasing $\langle n_{c}\rangle$ is due to 
the fact that decreasing $\langle n_{c}\rangle$ leads to the diminishing of the direct magnetic exchange on the $d-c$ bonds that scales 
like $V^2/(U_d+U_c)$ when $d$ and $c$-orbitals are both close to half-filling.
%As  $\langle n_{c}\rangle  $  is significantly decreased, only the Kondo coupling remains as the magnetic coupling between the localized $d$-electrons and $c$-electrons, which has an energy scale much smaller the direct exchanges, hence resulting the rapid decreasing of the $s_{\pm}$ wave pairing. 
See also below the fluctuation diagnostics on the pairing glue of $s_{\pm}$ wave.

%\ww{The remaining influence of $\langle n_{c}\rangle$ on low energy single-particle spectrum could only be, decreasing $\langle n_{c}\rangle$ brings an imbalance of $d-$ and $c-$weights between electron and hole pockets, which further impede effective inter-pocket pair scattering. From the perspective of pairing as a consequence of Kondo fluctuation, this imbalance essentially reflects insufficient screening of $d$ electrons with a strongly reduced Kondo scale known as Nozieres exhaustion citep{nozieres98,meyer00,pruschke00}.}

\subsection{Coexisting $d+is$ wave with interactions}

\begin{figure}
\includegraphics[scale=0.55]{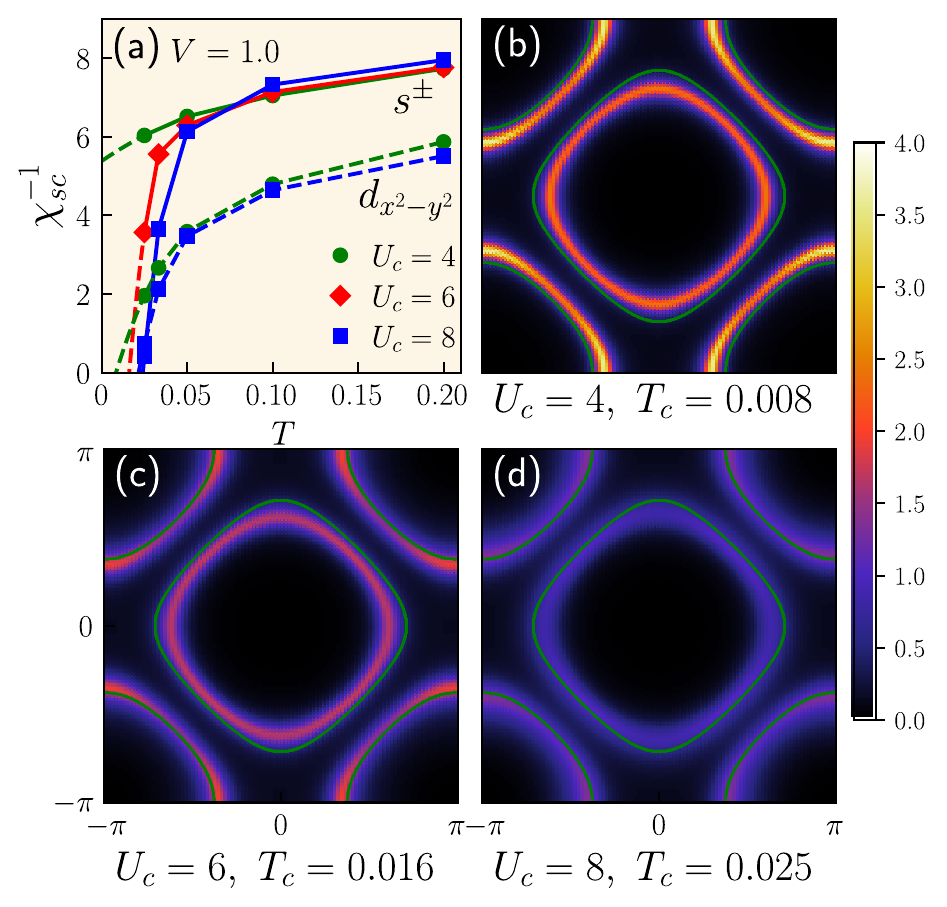}\caption{\label{fig:4}(a) The inverse pairing susceptibility $\chi_{sc}^{-1}$
as a function of $T$ with varying $U_{c}$ for $d$ (dashed lines)
and $s_{\pm}$ wave (solid lines) pairing. (b-d) FS with $U_{c}=4,6,8$.
For reference, the green lines indicate non-interacting FS. Here $V=1.0$,
$\langle n_{c}\rangle=0.9$ and $U_{c}=8$, $T=1/30$ are used.}
\end{figure}

In the above study, we found that $U_{c}=4$ is insufficient to stabilize a coexisting $d_{x^2-y^2}$ and $s_{\pm}$ wave pairing at different $\langle n_{c}\rangle$  and $V$. However,  by increasing $U_{c}$ up to 8, we did detect the coexistence of $d_{x^2-y^2}$ and $s_{\pm}$ wave pairing at the intermediate $V$.
%\out{In this regime our CDMFT reveals pronounced asymmetric feature on the self-energies, indicating breaking of $C_{4v}$ rotation symmetry \citep{kotliar88,nica20}.}
In order to determine the $T_{c}$ for both pairing symmetries, here we
separately compute the $d_{x^{2}-y^{2}}$ and $s_{\pm}$ wave pairing susceptibility for each parameter point
during the DMFT self-consistent loops.  In
Fig.~\ref{fig:4}(a), we plot $\chi_{sc}^{-1}$ as a
function of $T$ for both $d_{x^{2}-y^{2}}$ and $s_{\pm}$ wave at $V=1.0$.
One clearly sees that, as increasing $U_{c}$, $\chi_{sc}$
is drastically enhanced for $s_{\pm}$ wave at low $T$, while the
$d_{x^{2}-y^{2}}$ wave pairing susceptibility is affected more in a quantitative way. In particular, as $U_{c}=4$ is increased to $ U_{c}= 6$, a finite $T_c$ for $s_{\pm}$ wave pairing is reached, which signals the onset of the coexisting phase,
as  $T_c$ for $d_{x^{2}-y^{2}} $ wave is always finite  in this parameter regime (dashed lines).

%which should be associated with a great enhancement of inter-layer spin fluctuation.

Our results on the coexistence  of $d_{x^2-y^2}$ and $s_{\pm}$ wave for $V \in (0.8 - 1.1)$, $U_{c}=8$ can be summarized in the
driven of Fig.~\ref{fig:1},  which exhibits a phase diagram
 featuring the coexistence of $d_{x^2-y^2}$ and $s_{\pm}$  wave pairings. Comparing with $U_{c}=4$ result in the main plot, the $s_{\pm}$ wave instability in the driven at $U_{c}=8$  is enormously enhanced. For example, at $U_c = 4$, the minimal $V$ for  $s_{\pm}$ wave is around $V_{min} \sim 1.5$, while at $U_c = 8$ it is significantly reduced to $V_{min} \sim 0.85$.
For $d_{x^{2}-y^{2}}$ wave, the $T_{c}$ is also notably enhanced with increasing $U_c$,
for example, for $V=1.0,$ $T_{c} \approx 0.008  $ at $U_c=4$ (main plot)  while $T_c \approx 0.02$ at $U_c=8$ (Inset). To summarize, at $V\sim1$, $d_{x^{2}-y^{2}}$ and $s_{\pm}$ pairing have a comparable superconducting $T_c$, $T_c \approx 0.025$. As $V$ increases, $T_{c}$ of $s_{\pm}$ increases and that of  $d_{x^{2}-y^{2}}$ decreases, and otherwise the opposite.

When focusing on the FS evolution in Fig.~\ref{fig:4}(b-d), we found that such triggering effects on increasing $U_c$  is rarely reflected in FS geometry. Here, increasing $U_{c}$ only leads to a slight broadening of FS profile. This result also confirms our aforementioned argument that the role of FS is less important for $s_\pm$ wave pairing. %\ww{Note that the large $U_{c}$ is reminiscent of Anderson's doped  resonating valence bond (RVB) state in which inter-layer Kondo exchange serves as the pairing glue \citep{dongr13}.}

\subsection{Fluctuation diagnostics}

To unveil the driving forces for SC instabilities, here we apply fluctuation diagnostics \citep{Gunnarsson15,schafer21}.
The essential idea of this approach is to use the Schwinger-Dyson equation
of motion (SDEOM) to decompose the self-energies into a summation of  various two-particle
Green's functions. It has been recently used to reveal the significance of collective fluctuations
in the formation of $d$ wave superconductivity~\citep{Xinyang22}, pseudogap~\citep{Gunnarsson15},  and strange metal state~\citep{ww22}.
 Here we will focus on the fluctuation diagnostics
in the parameter regime of $d+is$ wave pairing. In the magnetic (spin) channel,
SDEOM of self-energy  can be expressed as
\begin{align}
&\left[\Sigma G\right]^{\textbf{k}}_{(m\alpha,n\beta)}=\nonumber\\
&-\frac{U_{m}}{N^{2}}\sum_{\textbf{k}^{\prime}\textbf{q}}G^{2}(\textbf{k}^{\prime}+\textbf{q}m\overset{-}{\alpha},\textbf{k}+\textbf{q}m\overset{-}{\alpha},\textbf{k}^{\prime}m\alpha,\textbf{k}n\beta),
\end{align}
in which $G^{2}(ijkl)=\langle\mathcal{T}c_{i}^{\dagger}c_{j}c_{k}c_{l}^{\dagger}\rangle$, 
$\textbf{k},\textbf{k}^{\prime}$ denote fermionic momenta/frequencies, and $\textbf{q}$ represents momentum transfer/bosonic frequency, $m,n$ denote orbital indices and $\alpha,\beta$ are spin indices.
Dividing the full Green's function matrix  $G$  on both sides of above equation, and sum over $\textbf{k}^{\prime}$,
we can obtain  the self-energy as the sum of four different fractions like
\begin{align}
\label{eq:sum}
\Sigma(\textbf{k})=\sum_{\textbf{q}}\left[\Sigma_{\textbf{q}}^{M,dd}+\Sigma_{\textbf{q}}^{M,dc}+\Sigma_{\textbf{q}}^{X,dd}+\Sigma_{\textbf{q}}^{X,dc}\right](\textbf{k}). 
\end{align}
Here $\Sigma_{\textbf{q}}^{M,mn}(\textbf{k})$ acquires a physical meaning that the relative weight of $\Sigma_{\textbf{q}}^{M,mn}(\textbf{k})$ at different $\mathbf{q}$  denotes  the importance of 
the magnetic fluctuation with transfer momentum $\textbf{q}$ in depleting electrons with momentum $\mathbf{k}$, since $\Sigma_{\textbf{q}}^{M,mn}  \propto  \langle S_{m}^{+}(\textbf{q})S_{n}^{-}(-\textbf{q})\rangle$~\citep{ww22},
while $\Sigma_{\textbf{q}}^{X,mn}$ is the fraction that beyond the description of
magnetic fluctuations, which can be alternatively seen as the fluctuation
between the spin and particle-particle (pairing) degrees of freedom: $\Sigma^{X,mn}_{\textbf{q}}\propto \langle \Delta_m(\textbf{q}) S_n^-(-\textbf{q}) \rangle$, with triplet pairing operator $\Delta_m(\textbf{q})=\sum_{\textbf{k}}(c_{m\textbf{k}\downarrow} c_{m\textbf{q}-\textbf{k}\downarrow})$.
In our study, we find that $\Sigma_{\textbf{q}}^{X,mn}$ are either negligible (for $d_{x^2-y^2}$ wave) or only have negative contributions (pair-breaking, for $s_{\pm}$ wave). Hence below we only discuss the $\Sigma_{\textbf{q}}^{M,mn}(\textbf{k})$ contributions.
In addition, here $dd$ or $dc$ denotes the contribution from the
fluctuations inside $d$-layer or between the $d-c$ layers. We stress that the
above sum rule in Eq.~(\ref{eq:sum}) of  fluctuation diagnostics is exact without any
approximation. 
%We will see it well reproduces the predominate antiferromagnetic fluctuation in single-band situation.

\begin{figure}[ht]
\includegraphics[scale=0.25]{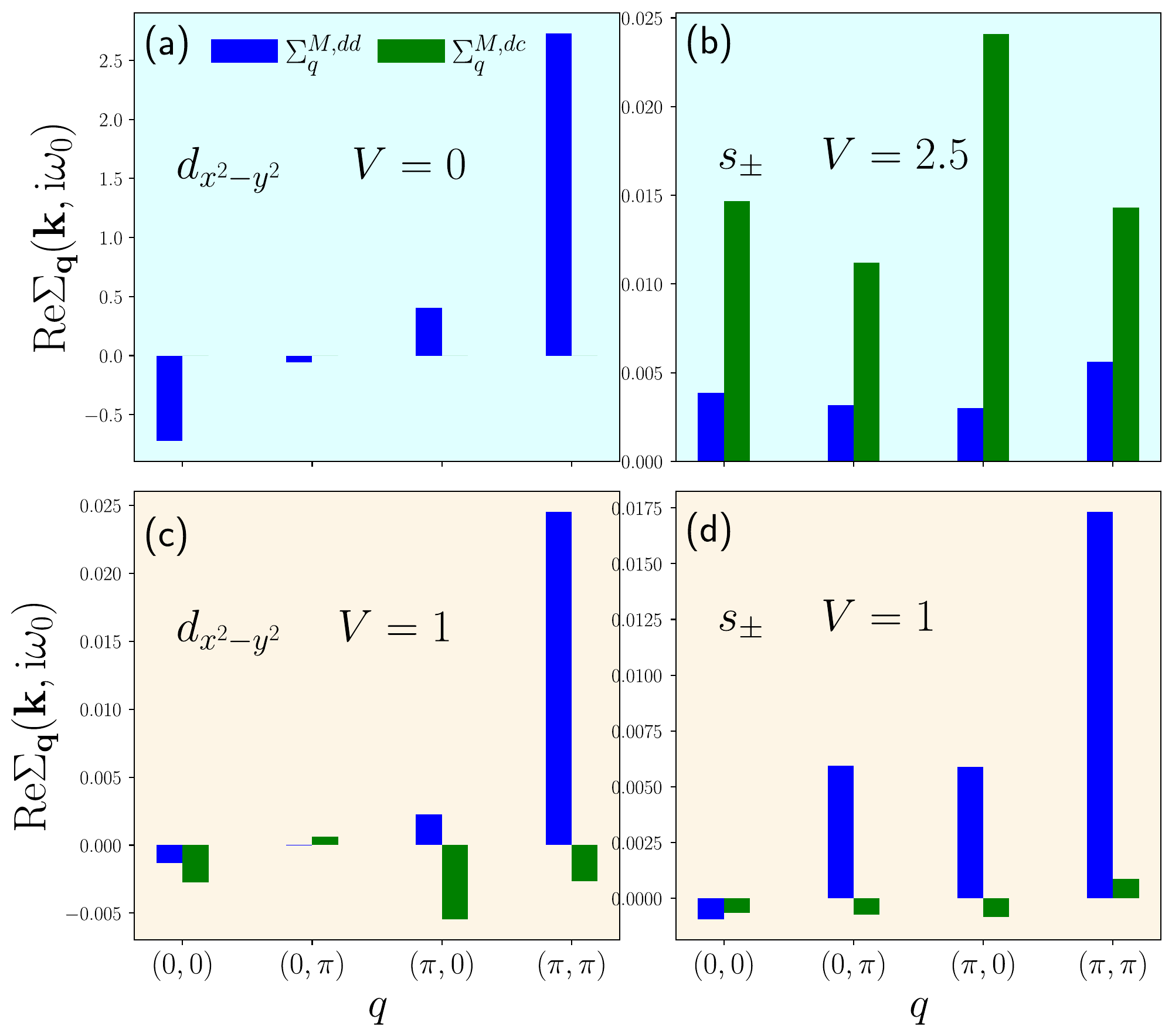}
\caption{\label{fig:5}(a) The fluctuation diagnostics of the anomalous self-energies
${\rm Re}\Sigma_{\textbf{q}}^{M,dd/dc}(\textbf{k},i\omega_{0})$
as a function of transfer momentum $\textbf{q}$ in the magnetic
channel. (a) $d_{x^{2}-y^{2}}$ wave pairing for $\textbf{k}=(0,\pi)$,
$V=0$. (b) $s_{\pm}$ wave pairing for $\textbf{k}=(0,0)$, $V=2.5$.
(c) $d_{x^{2}-y^{2}}$ and (d) $s_{\pm}$ wave pairing for $V=1,U_{c}=8$.
The temperature $T=1/30$.}
\end{figure}

Since we are most interested in the superconductivity, which is encoded in the real
part of anomalous self-energies. Focusing  on the low-energy physics,  we study the anomalous self-energies at the first Matsubara frequency
${\rm Re}\Sigma_{\textbf{q}}(\textbf{k},i\omega_{0})$. In
Fig.~\ref{fig:5}, we show the fluctuation diagnostics as a function of $\textbf{q}$,
with $T=1/30$. Fig.~\ref{fig:5}(a-b) respectively show $V=0$
and 2.5 with $U_{c}=4$. In Fig.~\ref{fig:5}(a), $\Sigma_{\textbf{q}}^{M,dc}$
(blue) has zero weight due to the decoupling of $d-c$ layers at $V=0$, in this case a predominated
$\Sigma_{(\pi,\pi)}^{M,dd}$ at $\textbf{q}=(\pi,\pi)$ overwhelms that of all others $q$, highlighting the
antiferromagnetic fluctuation origin of the  $d_{x^{2}-y^{2}}$ pairing  in a single band system~\citep{Xinyang22}.
In Fig.~\ref{fig:5}(b) where $s_\pm$ wave prevails, the major contributions
are from inter-layer $\Sigma_{\textbf{q}}^{M,dc}$ (green) which are quite evenly distributed among four different $\textbf{q}$, suggesting that here the anomalous self-energy, or the source of the pairing, should be seen as mainly stemming from the antiferromagnetic fluctuations between the  local  $d-c$ bond in real space.
% In this regime we also note another strong negative contribution from  $\Sigma_{\boldsymbol{q}}^{X,dc}$ (no shown), competiting with the former to determine an $s_{\pm}$ pairing in large $V$ regime. 
Fig.~\ref{fig:5}(c-d)
respectively display the result for  $d_{x^{2}-y^{2}}$ and $s_{\pm}$ wave pairing in
the $d+is$ coexistence regime at the intermediate $V$.  Comparing Fig.~\ref{fig:5}(c) with Fig.~\ref{fig:5}(a),
although the absolute magnitude of $\Sigma_{(\pi,\pi)}^{M,dd}$ (blue) is enormously depressed, the $\textbf{q}=(\pi,\pi)$ mode  retains its predominant role in anomalous scatterings,  which again emphasizes the vital role of antiferromagnetic fluctuations in $d_{x^2-y^2}$ wave pairing. 
However for Fig.~\ref{fig:5}(d), the situation is quite different as compared with Fig.~\ref{fig:5}(b). Here the fluctuation decomposition shows a major contribution from $\Sigma^{M,dd}_\textbf{q}$. In fact we find the magnitude of $\Sigma^{X,dd}_\textbf{q}$ is almost as strong as $\Sigma^{M,dd}_\textbf{q}$ but with the opposite signs (not shown here), which means they cancel out with each other ($\sum_{\textbf{q}}[\Sigma_{\textbf{q}}^{M,dd}+\Sigma_{\textbf{q}}^{X,dd}]\approx 0$). In this case another positive contribution from $\Sigma_{(\pi,\pi)}^{X,dc}$, although not as significant as the former two, is however not negligible for the $s_\pm$ wave pairing (no shown). These further suggest that the fluctuations of the $s_\pm$ wave in the coexistence regime is rather complicated and can not be simply attributed to a specific magnetic fluctuation channel based on the fluctuation diagnostics.

\section{\label{sec:5}Discussion}

In our systematic investigation, we have shown that for the  $d_{x^{2}-y^{2}}$ wave  pairing of  strongly correlated electrons, when a second  band with  a small $U$ is introduced,
the hybridization effects between the bands  will  effectively suppress the superconducting $T_c$. This may can 
be understand from two aspects. First, the hybridization  changes the geometry of the Fermi surface to energetically 
disfavor the $d_{x^{2}-y^{2}}$ wave form factor. Secondly,  the less correlated electrons from the second band can 
screen the localized electrons in the strongly correlated band, by suppressing the intra-band antiferromagnetic exchanges.
Therefore,  a multi-orbital description of nickelates involving the $5d$-orbitals  would inevitably lead to the
suppression of $d_{x^{2}-y^{2}}$ wave pairing~\citep{wuxx201}. In experiments, a recent resonant inelastic x-ray scattering study \citep{luh21}
reveals that the Ni-$3d$ electrons in the ground state of the undoped NdNiO\textsubscript{2} is an antiferromagnetically aligned with substantial damping, due to the coupling to rare-earth itinerant electrons.

Our study also demonstrates that  the coexistence of   $d_{x^{2}-y^{2}}$  and $s_{\pm}$ wave, or the  $d+is$ wave pairing is quite
elusive in a two-band system, which requires a large electron repulsion in the $c$-orbital and an intermediate hybridization $V\sim t$.
Regarding the nickelate materials, recent experimental evidences have shown that
the effective hybridization between Ni-$3d$ and Nd-derived orbitals is essentially small ~\citep{zhangh20,petocchi20,wuxx201,choi20,huangjw22}. Moreover, considering also
 the itinerant carrier density $\langle n_c\rangle$ is estimated to be only 0.08 per
unit cell \citep{Lidanfeng_PRL20}, which is one order of magnitude smaller than our estimation for a finite $T_c$ for $s_{\pm}$ wave pairing. Therefore, we conjecture that the potential $s_{\pm}$ wave, as well as the coexisting $d+is$ wave pairing, would have an extremely small $T_c$, if it ever exists in  infinite layer nickelates. 
We note that in the tunneling experiment \cite{guqq20}, the $d_{x^2-y^2}$ V-shaped gap is reported to be an intrinsic feature in the  infinite layer nickelates, while U-shaped  $s_\pm$ wave gap seems more associated with a rough surface. In London penetration experiments, $d+is$ wave pairing was also reported~\cite{chow22}, where the full gap, however,  was also argued to be attributed  to the magnetic impurities~\cite{harvey22}.

%Finally, we also notice a recent DFT+DMFT
%study that emphasizes the role of Nozieres exhaustion {[https://meetings.aps.org/Meeting/MAR23/Session/A24.3]}. We look
%forward to further experimental evidences as well as theoretical interpretations
%on SC pairing symmetry in nickelates.

\section{\label{sec:6}Conclusion}

In summary, we have established a  phase diagram of $d_{x^{2}-y^{2}}$, $s_{\pm}$
and $d+is$ wave pairing in a two-orbital Hubbard model, in which we clarify  the roles of
hybridization ($V$), itinerant carrier density ($\langle n_{c}\rangle$)
and Hubbard interaction ($U_{c}$) in determining the superconducting $T_c$. Our study have demonstrated that
the entrance of the $5d$-electrons at Fermi level hazards the $d_{x^{2}-y^{2}}$ wave pairing 
in the Ni-$3d$  orbital. On the other hand, the two-orbital $d+is$ wave pairing, however, requires a relative large Coulomb repulsion in the  $5d$-orbital.
We have also performed the fluctuation diagnostics to reveal the driving force behind the $d_{x^{2}-y^{2}}$ and $s_{\pm}$ wave pairing in this system. Our study overall supports an intrinsic $d_{x^{2}-y^{2}}$ wave pairing in the nickelate superconductors despite the contamination of rare-earth derived bands at the Fermi level.

\begin{acknowledgments}
We are grateful to Mi Jiang for useful discussions.
This project was supported by the National
Key Research and Development Program of China
(Grant No. 2022YFA1402802), the
National Natural Science Foundation of China (Grants
No. 12274472, No. 92165204, No. 11974432,
), Shenzhen International
Quantum Academy (Grant No. SIQA202102), and GuangZhou National supercomputing center.
\end{acknowledgments}

\bibliography{Ni}

\begin{thebibliography}{69}%
\makeatletter
\providecommand \@ifxundefined [1]{%
 \@ifx{#1\undefined}
}%
\providecommand \@ifnum [1]{%
 \ifnum #1\expandafter \@firstoftwo
 \else \expandafter \@secondoftwo
 \fi
}%
\providecommand \@ifx [1]{%
 \ifx #1\expandafter \@firstoftwo
 \else \expandafter \@secondoftwo
 \fi
}%
\providecommand \natexlab [1]{#1}%
\providecommand \enquote  [1]{``#1''}%
\providecommand \bibnamefont  [1]{#1}%
\providecommand \bibfnamefont [1]{#1}%
\providecommand \citenamefont [1]{#1}%
\providecommand \href@noop [0]{\@secondoftwo}%
\providecommand \href [0]{\begingroup \@sanitize@url \@href}%
\providecommand \@href[1]{\@@startlink{#1}\@@href}%
\providecommand \@@href[1]{\endgroup#1\@@endlink}%
\providecommand \@sanitize@url [0]{\catcode `\\12\catcode `\$12\catcode `\&12\catcode `\#12\catcode `\^12\catcode `\_12\catcode `\%12\relax}%
\providecommand \@@startlink[1]{}%
\providecommand \@@endlink[0]{}%
\providecommand \url  [0]{\begingroup\@sanitize@url \@url }%
\providecommand \@url [1]{\endgroup\@href {#1}{\urlprefix }}%
\providecommand \urlprefix  [0]{URL }%
\providecommand \Eprint [0]{\href }%
\providecommand \doibase [0]{http://dx.doi.org/}%
\providecommand \selectlanguage [0]{\@gobble}%
\providecommand \bibinfo  [0]{\@secondoftwo}%
\providecommand \bibfield  [0]{\@secondoftwo}%
\providecommand \translation [1]{[#1]}%
\providecommand \BibitemOpen [0]{}%
\providecommand \bibitemStop [0]{}%
\providecommand \bibitemNoStop [0]{.\EOS\space}%
\providecommand \EOS [0]{\spacefactor3000\relax}%
\providecommand \BibitemShut  [1]{\csname bibitem#1\endcsname}%
\let\auto@bib@innerbib\@empty
%</preamble>
\bibitem [{\citenamefont {Lee}\ \emph {et~al.}(2006)\citenamefont {Lee}, \citenamefont {Nagaosa},\ and\ \citenamefont {Wen}}]{patrick06}%
  \BibitemOpen
  \bibfield  {author} {\bibinfo {author} {\bibfnamefont {P.~A.}\ \bibnamefont {Lee}}, \bibinfo {author} {\bibfnamefont {N.}~\bibnamefont {Nagaosa}}, \ and\ \bibinfo {author} {\bibfnamefont {X.-G.}\ \bibnamefont {Wen}},\ }\href {\doibase 10.1103/RevModPhys.78.17} {\bibfield  {journal} {\bibinfo  {journal} {Reviews of Modern Physics}\ }\textbf {\bibinfo {volume} {78}},\ \bibinfo {pages} {17} (\bibinfo {year} {2006})}\BibitemShut {NoStop}%
\bibitem [{\citenamefont {Lee}\ \emph {et~al.}(2009)\citenamefont {Lee}, \citenamefont {Zhang},\ and\ \citenamefont {Wu}}]{leewc09}%
  \BibitemOpen
  \bibfield  {author} {\bibinfo {author} {\bibfnamefont {W.~C.}\ \bibnamefont {Lee}}, \bibinfo {author} {\bibfnamefont {S.~C.}\ \bibnamefont {Zhang}}, \ and\ \bibinfo {author} {\bibfnamefont {C.}~\bibnamefont {Wu}},\ }\href {\doibase 10.1103/PhysRevLett.102.217002} {\bibfield  {journal} {\bibinfo  {journal} {Phys Rev Lett}\ }\textbf {\bibinfo {volume} {102}},\ \bibinfo {pages} {217002} (\bibinfo {year} {2009})}\BibitemShut {NoStop}%
\bibitem [{\citenamefont {Fernandes}\ and\ \citenamefont {Millis}(2013)}]{Fernandes13}%
  \BibitemOpen
  \bibfield  {author} {\bibinfo {author} {\bibfnamefont {R.~M.}\ \bibnamefont {Fernandes}}\ and\ \bibinfo {author} {\bibfnamefont {A.~J.}\ \bibnamefont {Millis}},\ }\href {\doibase 10.1103/physrevlett.111.127001} {\bibfield  {journal} {\bibinfo  {journal} {Physical Review Letters}\ }\textbf {\bibinfo {volume} {111}} (\bibinfo {year} {2013}),\ 10.1103/physrevlett.111.127001}\BibitemShut {NoStop}%
\bibitem [{\citenamefont {Platt}\ \emph {et~al.}(2012)\citenamefont {Platt}, \citenamefont {Thomale}, \citenamefont {Honerkamp}, \citenamefont {Zhang},\ and\ \citenamefont {Hanke}}]{platt12}%
  \BibitemOpen
  \bibfield  {author} {\bibinfo {author} {\bibfnamefont {C.}~\bibnamefont {Platt}}, \bibinfo {author} {\bibfnamefont {R.}~\bibnamefont {Thomale}}, \bibinfo {author} {\bibfnamefont {C.}~\bibnamefont {Honerkamp}}, \bibinfo {author} {\bibfnamefont {S.-C.}\ \bibnamefont {Zhang}}, \ and\ \bibinfo {author} {\bibfnamefont {W.}~\bibnamefont {Hanke}},\ }\href {\doibase 10.1103/PhysRevB.85.180502} {\bibfield  {journal} {\bibinfo  {journal} {Physical Review B}\ }\textbf {\bibinfo {volume} {85}} (\bibinfo {year} {2012}),\ 10.1103/PhysRevB.85.180502}\BibitemShut {NoStop}%
\bibitem [{\citenamefont {Maier}\ \emph {et~al.}(2011)\citenamefont {Maier}, \citenamefont {Graser}, \citenamefont {Hirschfeld},\ and\ \citenamefont {Scalapino}}]{maier11_d}%
  \BibitemOpen
  \bibfield  {author} {\bibinfo {author} {\bibfnamefont {T.~A.}\ \bibnamefont {Maier}}, \bibinfo {author} {\bibfnamefont {S.}~\bibnamefont {Graser}}, \bibinfo {author} {\bibfnamefont {P.~J.}\ \bibnamefont {Hirschfeld}}, \ and\ \bibinfo {author} {\bibfnamefont {D.~J.}\ \bibnamefont {Scalapino}},\ }\href {\doibase 10.1103/PhysRevB.83.100515} {\bibfield  {journal} {\bibinfo  {journal} {Physical Review B}\ }\textbf {\bibinfo {volume} {83}} (\bibinfo {year} {2011}),\ 10.1103/PhysRevB.83.100515}\BibitemShut {NoStop}%
\bibitem [{\citenamefont {Li}\ \emph {et~al.}(2019)\citenamefont {Li}, \citenamefont {Lee}, \citenamefont {Wang}, \citenamefont {Osada}, \citenamefont {Crossley}, \citenamefont {Lee}, \citenamefont {Cui}, \citenamefont {Hikita},\ and\ \citenamefont {Hwang}}]{Lidanfeng_nature19}%
  \BibitemOpen
  \bibfield  {author} {\bibinfo {author} {\bibfnamefont {D.}~\bibnamefont {Li}}, \bibinfo {author} {\bibfnamefont {K.}~\bibnamefont {Lee}}, \bibinfo {author} {\bibfnamefont {B.~Y.}\ \bibnamefont {Wang}}, \bibinfo {author} {\bibfnamefont {M.}~\bibnamefont {Osada}}, \bibinfo {author} {\bibfnamefont {S.}~\bibnamefont {Crossley}}, \bibinfo {author} {\bibfnamefont {H.~R.}\ \bibnamefont {Lee}}, \bibinfo {author} {\bibfnamefont {Y.}~\bibnamefont {Cui}}, \bibinfo {author} {\bibfnamefont {Y.}~\bibnamefont {Hikita}}, \ and\ \bibinfo {author} {\bibfnamefont {H.~Y.}\ \bibnamefont {Hwang}},\ }\href {\doibase 10.1038/s41586-019-1496-5} {\bibfield  {journal} {\bibinfo  {journal} {Nature}\ }\textbf {\bibinfo {volume} {572}},\ \bibinfo {pages} {624} (\bibinfo {year} {2019})}\BibitemShut {NoStop}%
\bibitem [{\citenamefont {Li}\ \emph {et~al.}(2020)\citenamefont {Li}, \citenamefont {Wang}, \citenamefont {Lee}, \citenamefont {Harvey}, \citenamefont {Osada}, \citenamefont {Goodge}, \citenamefont {Kourkoutis},\ and\ \citenamefont {Hwang}}]{Lidanfeng_PRL20}%
  \BibitemOpen
  \bibfield  {author} {\bibinfo {author} {\bibfnamefont {D.}~\bibnamefont {Li}}, \bibinfo {author} {\bibfnamefont {B.~Y.}\ \bibnamefont {Wang}}, \bibinfo {author} {\bibfnamefont {K.}~\bibnamefont {Lee}}, \bibinfo {author} {\bibfnamefont {S.~P.}\ \bibnamefont {Harvey}}, \bibinfo {author} {\bibfnamefont {M.}~\bibnamefont {Osada}}, \bibinfo {author} {\bibfnamefont {B.~H.}\ \bibnamefont {Goodge}}, \bibinfo {author} {\bibfnamefont {L.~F.}\ \bibnamefont {Kourkoutis}}, \ and\ \bibinfo {author} {\bibfnamefont {H.~Y.}\ \bibnamefont {Hwang}},\ }\href {\doibase 10.1103/physrevlett.125.027001} {\bibfield  {journal} {\bibinfo  {journal} {Physical Review Letters}\ }\textbf {\bibinfo {volume} {125}} (\bibinfo {year} {2020}),\ 10.1103/physrevlett.125.027001}\BibitemShut {NoStop}%
\bibitem [{\citenamefont {Zeng}\ \emph {et~al.}(2020)\citenamefont {Zeng}, \citenamefont {Tang}, \citenamefont {Yin}, \citenamefont {Li}, \citenamefont {Li}, \citenamefont {Huang}, \citenamefont {Hu}, \citenamefont {Liu}, \citenamefont {Omar}, \citenamefont {Jani}, \citenamefont {Lim}, \citenamefont {Han}, \citenamefont {Wan}, \citenamefont {Yang}, \citenamefont {Pennycook}, \citenamefont {Wee},\ and\ \citenamefont {Ariando}}]{ZengShengwei_PRL20}%
  \BibitemOpen
  \bibfield  {author} {\bibinfo {author} {\bibfnamefont {S.}~\bibnamefont {Zeng}}, \bibinfo {author} {\bibfnamefont {C.~S.}\ \bibnamefont {Tang}}, \bibinfo {author} {\bibfnamefont {X.}~\bibnamefont {Yin}}, \bibinfo {author} {\bibfnamefont {C.}~\bibnamefont {Li}}, \bibinfo {author} {\bibfnamefont {M.}~\bibnamefont {Li}}, \bibinfo {author} {\bibfnamefont {Z.}~\bibnamefont {Huang}}, \bibinfo {author} {\bibfnamefont {J.}~\bibnamefont {Hu}}, \bibinfo {author} {\bibfnamefont {W.}~\bibnamefont {Liu}}, \bibinfo {author} {\bibfnamefont {G.~J.}\ \bibnamefont {Omar}}, \bibinfo {author} {\bibfnamefont {H.}~\bibnamefont {Jani}}, \bibinfo {author} {\bibfnamefont {Z.~S.}\ \bibnamefont {Lim}}, \bibinfo {author} {\bibfnamefont {K.}~\bibnamefont {Han}}, \bibinfo {author} {\bibfnamefont {D.}~\bibnamefont {Wan}}, \bibinfo {author} {\bibfnamefont {P.}~\bibnamefont {Yang}}, \bibinfo {author} {\bibfnamefont {S.~J.}\ \bibnamefont {Pennycook}}, \bibinfo {author} {\bibfnamefont {A.~T.}\ \bibnamefont {Wee}}, \ and\ \bibinfo {author}
  {\bibfnamefont {A.}~\bibnamefont {Ariando}},\ }\href {\doibase 10.1103/physrevlett.125.147003} {\bibfield  {journal} {\bibinfo  {journal} {Physical Review Letters}\ }\textbf {\bibinfo {volume} {125}} (\bibinfo {year} {2020}),\ 10.1103/physrevlett.125.147003}\BibitemShut {NoStop}%
\bibitem [{\citenamefont {Gu}\ \emph {et~al.}(2020)\citenamefont {Gu}, \citenamefont {Li}, \citenamefont {Wan}, \citenamefont {Li}, \citenamefont {Guo}, \citenamefont {Yang}, \citenamefont {Li}, \citenamefont {Zhu}, \citenamefont {Pan}, \citenamefont {Nie},\ and\ \citenamefont {Wen}}]{guqq20}%
  \BibitemOpen
  \bibfield  {author} {\bibinfo {author} {\bibfnamefont {Q.}~\bibnamefont {Gu}}, \bibinfo {author} {\bibfnamefont {Y.}~\bibnamefont {Li}}, \bibinfo {author} {\bibfnamefont {S.}~\bibnamefont {Wan}}, \bibinfo {author} {\bibfnamefont {H.}~\bibnamefont {Li}}, \bibinfo {author} {\bibfnamefont {W.}~\bibnamefont {Guo}}, \bibinfo {author} {\bibfnamefont {H.}~\bibnamefont {Yang}}, \bibinfo {author} {\bibfnamefont {Q.}~\bibnamefont {Li}}, \bibinfo {author} {\bibfnamefont {X.}~\bibnamefont {Zhu}}, \bibinfo {author} {\bibfnamefont {X.}~\bibnamefont {Pan}}, \bibinfo {author} {\bibfnamefont {Y.}~\bibnamefont {Nie}}, \ and\ \bibinfo {author} {\bibfnamefont {H.-H.}\ \bibnamefont {Wen}},\ }\href {\doibase 10.1038/s41467-020-19908-1} {\bibfield  {journal} {\bibinfo  {journal} {Nature Communications}\ }\textbf {\bibinfo {volume} {11}} (\bibinfo {year} {2020}),\ 10.1038/s41467-020-19908-1}\BibitemShut {NoStop}%
\bibitem [{\citenamefont {Osada}\ \emph {et~al.}(2020{\natexlab{a}})\citenamefont {Osada}, \citenamefont {Wang}, \citenamefont {Goodge}, \citenamefont {Lee}, \citenamefont {Yoon}, \citenamefont {Sakuma}, \citenamefont {Li}, \citenamefont {Miura}, \citenamefont {Kourkoutis},\ and\ \citenamefont {Hwang}}]{oasda20}%
  \BibitemOpen
  \bibfield  {author} {\bibinfo {author} {\bibfnamefont {M.}~\bibnamefont {Osada}}, \bibinfo {author} {\bibfnamefont {B.~Y.}\ \bibnamefont {Wang}}, \bibinfo {author} {\bibfnamefont {B.~H.}\ \bibnamefont {Goodge}}, \bibinfo {author} {\bibfnamefont {K.}~\bibnamefont {Lee}}, \bibinfo {author} {\bibfnamefont {H.}~\bibnamefont {Yoon}}, \bibinfo {author} {\bibfnamefont {K.}~\bibnamefont {Sakuma}}, \bibinfo {author} {\bibfnamefont {D.}~\bibnamefont {Li}}, \bibinfo {author} {\bibfnamefont {M.}~\bibnamefont {Miura}}, \bibinfo {author} {\bibfnamefont {L.~F.}\ \bibnamefont {Kourkoutis}}, \ and\ \bibinfo {author} {\bibfnamefont {H.~Y.}\ \bibnamefont {Hwang}},\ }\href {\doibase 10.1021/acs.nanolett.0c01392} {\bibfield  {journal} {\bibinfo  {journal} {Nano Letters}\ }\textbf {\bibinfo {volume} {20}},\ \bibinfo {pages} {5735} (\bibinfo {year} {2020}{\natexlab{a}})}\BibitemShut {NoStop}%
\bibitem [{\citenamefont {Osada}\ \emph {et~al.}(2020{\natexlab{b}})\citenamefont {Osada}, \citenamefont {Wang}, \citenamefont {Lee}, \citenamefont {Li},\ and\ \citenamefont {Hwang}}]{osada20_prm}%
  \BibitemOpen
  \bibfield  {author} {\bibinfo {author} {\bibfnamefont {M.}~\bibnamefont {Osada}}, \bibinfo {author} {\bibfnamefont {B.~Y.}\ \bibnamefont {Wang}}, \bibinfo {author} {\bibfnamefont {K.}~\bibnamefont {Lee}}, \bibinfo {author} {\bibfnamefont {D.}~\bibnamefont {Li}}, \ and\ \bibinfo {author} {\bibfnamefont {H.~Y.}\ \bibnamefont {Hwang}},\ }\href {\doibase 10.1103/PhysRevMaterials.4.121801} {\bibfield  {journal} {\bibinfo  {journal} {Physical Review Materials}\ }\textbf {\bibinfo {volume} {4}} (\bibinfo {year} {2020}{\natexlab{b}}),\ 10.1103/PhysRevMaterials.4.121801}\BibitemShut {NoStop}%
\bibitem [{\citenamefont {Zhou}\ \emph {et~al.}(2021)\citenamefont {Zhou}, \citenamefont {Feng}, \citenamefont {Qin}, \citenamefont {Yan}, \citenamefont {Wang}, \citenamefont {Nie}, \citenamefont {Wu}, \citenamefont {Zhang}, \citenamefont {Chen}, \citenamefont {Meng}, \citenamefont {Zhu},\ and\ \citenamefont {Liu}}]{zhouxr21}%
  \BibitemOpen
  \bibfield  {author} {\bibinfo {author} {\bibfnamefont {X.-R.}\ \bibnamefont {Zhou}}, \bibinfo {author} {\bibfnamefont {Z.-X.}\ \bibnamefont {Feng}}, \bibinfo {author} {\bibfnamefont {P.-X.}\ \bibnamefont {Qin}}, \bibinfo {author} {\bibfnamefont {H.}~\bibnamefont {Yan}}, \bibinfo {author} {\bibfnamefont {X.-N.}\ \bibnamefont {Wang}}, \bibinfo {author} {\bibfnamefont {P.}~\bibnamefont {Nie}}, \bibinfo {author} {\bibfnamefont {H.-J.}\ \bibnamefont {Wu}}, \bibinfo {author} {\bibfnamefont {X.}~\bibnamefont {Zhang}}, \bibinfo {author} {\bibfnamefont {H.-Y.}\ \bibnamefont {Chen}}, \bibinfo {author} {\bibfnamefont {Z.-A.}\ \bibnamefont {Meng}}, \bibinfo {author} {\bibfnamefont {Z.-W.}\ \bibnamefont {Zhu}}, \ and\ \bibinfo {author} {\bibfnamefont {Z.-Q.}\ \bibnamefont {Liu}},\ }\href {\doibase 10.1007/s12598-021-01768-3} {\bibfield  {journal} {\bibinfo  {journal} {Rare Metals}\ } (\bibinfo {year} {2021}),\ 10.1007/s12598-021-01768-3}\BibitemShut {NoStop}%
\bibitem [{\citenamefont {Pan}\ \emph {et~al.}(2021)\citenamefont {Pan}, \citenamefont {Segedin}, \citenamefont {LaBollita}, \citenamefont {Song}, \citenamefont {Nica}, \citenamefont {Goodge}, \citenamefont {Pierce}, \citenamefont {Doyle}, \citenamefont {Novakov}, \citenamefont {Carrizales}, \citenamefont {N’Diaye}, \citenamefont {Shafer}, \citenamefont {Paik}, \citenamefont {Heron}, \citenamefont {Mason}, \citenamefont {Yacoby}, \citenamefont {Kourkoutis}, \citenamefont {Erten}, \citenamefont {Brooks}, \citenamefont {Botana},\ and\ \citenamefont {Mundy}}]{pangrace21}%
  \BibitemOpen
  \bibfield  {author} {\bibinfo {author} {\bibfnamefont {G.~A.}\ \bibnamefont {Pan}}, \bibinfo {author} {\bibfnamefont {D.~F.}\ \bibnamefont {Segedin}}, \bibinfo {author} {\bibfnamefont {H.}~\bibnamefont {LaBollita}}, \bibinfo {author} {\bibfnamefont {Q.}~\bibnamefont {Song}}, \bibinfo {author} {\bibfnamefont {E.~M.}\ \bibnamefont {Nica}}, \bibinfo {author} {\bibfnamefont {B.~H.}\ \bibnamefont {Goodge}}, \bibinfo {author} {\bibfnamefont {A.~T.}\ \bibnamefont {Pierce}}, \bibinfo {author} {\bibfnamefont {S.}~\bibnamefont {Doyle}}, \bibinfo {author} {\bibfnamefont {S.}~\bibnamefont {Novakov}}, \bibinfo {author} {\bibfnamefont {D.~C.}\ \bibnamefont {Carrizales}}, \bibinfo {author} {\bibfnamefont {A.~T.}\ \bibnamefont {N’Diaye}}, \bibinfo {author} {\bibfnamefont {P.}~\bibnamefont {Shafer}}, \bibinfo {author} {\bibfnamefont {H.}~\bibnamefont {Paik}}, \bibinfo {author} {\bibfnamefont {J.~T.}\ \bibnamefont {Heron}}, \bibinfo {author} {\bibfnamefont {J.~A.}\ \bibnamefont {Mason}}, \bibinfo {author} {\bibfnamefont
  {A.}~\bibnamefont {Yacoby}}, \bibinfo {author} {\bibfnamefont {L.~F.}\ \bibnamefont {Kourkoutis}}, \bibinfo {author} {\bibfnamefont {O.}~\bibnamefont {Erten}}, \bibinfo {author} {\bibfnamefont {C.~M.}\ \bibnamefont {Brooks}}, \bibinfo {author} {\bibfnamefont {A.~S.}\ \bibnamefont {Botana}}, \ and\ \bibinfo {author} {\bibfnamefont {J.~A.}\ \bibnamefont {Mundy}},\ }\href@noop {} {\bibfield  {journal} {\bibinfo  {journal} {Nature Materials}\ } (\bibinfo {year} {2021})}\BibitemShut {NoStop}%
\bibitem [{\citenamefont {Osada}\ \emph {et~al.}(2021)\citenamefont {Osada}, \citenamefont {Wang}, \citenamefont {Goodge}, \citenamefont {Harvey}, \citenamefont {Lee}, \citenamefont {Li}, \citenamefont {Kourkoutis},\ and\ \citenamefont {Hwang}}]{osada21}%
  \BibitemOpen
  \bibfield  {author} {\bibinfo {author} {\bibfnamefont {M.}~\bibnamefont {Osada}}, \bibinfo {author} {\bibfnamefont {B.~Y.}\ \bibnamefont {Wang}}, \bibinfo {author} {\bibfnamefont {B.~H.}\ \bibnamefont {Goodge}}, \bibinfo {author} {\bibfnamefont {S.~P.}\ \bibnamefont {Harvey}}, \bibinfo {author} {\bibfnamefont {K.}~\bibnamefont {Lee}}, \bibinfo {author} {\bibfnamefont {D.}~\bibnamefont {Li}}, \bibinfo {author} {\bibfnamefont {L.~F.}\ \bibnamefont {Kourkoutis}}, \ and\ \bibinfo {author} {\bibfnamefont {H.~Y.}\ \bibnamefont {Hwang}},\ }\href {\doibase 10.1002/adma.202104083} {\bibfield  {journal} {\bibinfo  {journal} {Adv Mater}\ }\textbf {\bibinfo {volume} {33}},\ \bibinfo {pages} {e2104083} (\bibinfo {year} {2021})}\BibitemShut {NoStop}%
\bibitem [{\citenamefont {Zeng}\ \emph {et~al.}(2022{\natexlab{a}})\citenamefont {Zeng}, \citenamefont {Li}, \citenamefont {Chow}, \citenamefont {Cao}, \citenamefont {Zhang}, \citenamefont {Tang}, \citenamefont {Yin}, \citenamefont {Lim}, \citenamefont {Hu}, \citenamefont {Yang},\ and\ \citenamefont {Ariando}}]{shengwei22}%
  \BibitemOpen
  \bibfield  {author} {\bibinfo {author} {\bibfnamefont {S.}~\bibnamefont {Zeng}}, \bibinfo {author} {\bibfnamefont {C.}~\bibnamefont {Li}}, \bibinfo {author} {\bibfnamefont {L.~E.}\ \bibnamefont {Chow}}, \bibinfo {author} {\bibfnamefont {Y.}~\bibnamefont {Cao}}, \bibinfo {author} {\bibfnamefont {Z.}~\bibnamefont {Zhang}}, \bibinfo {author} {\bibfnamefont {C.~S.}\ \bibnamefont {Tang}}, \bibinfo {author} {\bibfnamefont {X.}~\bibnamefont {Yin}}, \bibinfo {author} {\bibfnamefont {Z.~S.}\ \bibnamefont {Lim}}, \bibinfo {author} {\bibfnamefont {J.}~\bibnamefont {Hu}}, \bibinfo {author} {\bibfnamefont {P.}~\bibnamefont {Yang}}, \ and\ \bibinfo {author} {\bibfnamefont {A.}~\bibnamefont {Ariando}},\ }\href {\doibase 10.1126/sciadv.abl9927} {\bibfield  {journal} {\bibinfo  {journal} {Science Advances}\ }\textbf {\bibinfo {volume} {8}},\ \bibinfo {pages} {eabl9927} (\bibinfo {year} {2022}{\natexlab{a}})},\ \Eprint {http://arxiv.org/abs/https://www.science.org/doi/pdf/10.1126/sciadv.abl9927}
  {https://www.science.org/doi/pdf/10.1126/sciadv.abl9927} \BibitemShut {NoStop}%
\bibitem [{\citenamefont {Zeng}\ \emph {et~al.}(2022{\natexlab{b}})\citenamefont {Zeng}, \citenamefont {Yin}, \citenamefont {Li}, \citenamefont {Chow}, \citenamefont {Tang}, \citenamefont {Han}, \citenamefont {Huang}, \citenamefont {Cao}, \citenamefont {Wan}, \citenamefont {Zhang}, \citenamefont {Lim}, \citenamefont {Diao}, \citenamefont {Yang}, \citenamefont {Wee}, \citenamefont {Pennycook},\ and\ \citenamefont {Ariando}}]{zengsw22}%
  \BibitemOpen
  \bibfield  {author} {\bibinfo {author} {\bibfnamefont {S.~W.}\ \bibnamefont {Zeng}}, \bibinfo {author} {\bibfnamefont {X.~M.}\ \bibnamefont {Yin}}, \bibinfo {author} {\bibfnamefont {C.~J.}\ \bibnamefont {Li}}, \bibinfo {author} {\bibfnamefont {L.~E.}\ \bibnamefont {Chow}}, \bibinfo {author} {\bibfnamefont {C.~S.}\ \bibnamefont {Tang}}, \bibinfo {author} {\bibfnamefont {K.}~\bibnamefont {Han}}, \bibinfo {author} {\bibfnamefont {Z.}~\bibnamefont {Huang}}, \bibinfo {author} {\bibfnamefont {Y.}~\bibnamefont {Cao}}, \bibinfo {author} {\bibfnamefont {D.~Y.}\ \bibnamefont {Wan}}, \bibinfo {author} {\bibfnamefont {Z.~T.}\ \bibnamefont {Zhang}}, \bibinfo {author} {\bibfnamefont {Z.~S.}\ \bibnamefont {Lim}}, \bibinfo {author} {\bibfnamefont {C.~Z.}\ \bibnamefont {Diao}}, \bibinfo {author} {\bibfnamefont {P.}~\bibnamefont {Yang}}, \bibinfo {author} {\bibfnamefont {A.~T.~S.}\ \bibnamefont {Wee}}, \bibinfo {author} {\bibfnamefont {S.~J.}\ \bibnamefont {Pennycook}}, \ and\ \bibinfo {author} {\bibfnamefont
  {A.}~\bibnamefont {Ariando}},\ }\href {\doibase 10.1038/s41467-022-28390-w} {\bibfield  {journal} {\bibinfo  {journal} {Nat Commun}\ }\textbf {\bibinfo {volume} {13}},\ \bibinfo {pages} {743} (\bibinfo {year} {2022}{\natexlab{b}})}\BibitemShut {NoStop}%
\bibitem [{\citenamefont {Chow}\ \emph {et~al.}(2023)\citenamefont {Chow}, \citenamefont {Sudheesh}, \citenamefont {Luo}, \citenamefont {Nandi}, \citenamefont {Heil}, \citenamefont {Deuschle}, \citenamefont {Zeng}, \citenamefont {Zhang}, \citenamefont {Prakash}, \citenamefont {Du}, \citenamefont {Lim}, \citenamefont {van Aken}, \citenamefont {Chia},\ and\ \citenamefont {Ariando}}]{chow22}%
  \BibitemOpen
  \bibfield  {author} {\bibinfo {author} {\bibfnamefont {L.~E.}\ \bibnamefont {Chow}}, \bibinfo {author} {\bibfnamefont {S.~K.}\ \bibnamefont {Sudheesh}}, \bibinfo {author} {\bibfnamefont {Z.~Y.}\ \bibnamefont {Luo}}, \bibinfo {author} {\bibfnamefont {P.}~\bibnamefont {Nandi}}, \bibinfo {author} {\bibfnamefont {T.}~\bibnamefont {Heil}}, \bibinfo {author} {\bibfnamefont {J.}~\bibnamefont {Deuschle}}, \bibinfo {author} {\bibfnamefont {S.~W.}\ \bibnamefont {Zeng}}, \bibinfo {author} {\bibfnamefont {Z.~T.}\ \bibnamefont {Zhang}}, \bibinfo {author} {\bibfnamefont {S.}~\bibnamefont {Prakash}}, \bibinfo {author} {\bibfnamefont {X.~M.}\ \bibnamefont {Du}}, \bibinfo {author} {\bibfnamefont {Z.~S.}\ \bibnamefont {Lim}}, \bibinfo {author} {\bibfnamefont {P.~A.}\ \bibnamefont {van Aken}}, \bibinfo {author} {\bibfnamefont {E.~E.~M.}\ \bibnamefont {Chia}}, \ and\ \bibinfo {author} {\bibfnamefont {A.}~\bibnamefont {Ariando}},\ }\href@noop {} {\enquote {\bibinfo {title} {Pairing symmetry in infinite-layer nickelate
  superconductor},}\ } (\bibinfo {year} {2023}),\ \Eprint {http://arxiv.org/abs/2201.10038} {arXiv:2201.10038 [cond-mat.supr-con]} \BibitemShut {NoStop}%
\bibitem [{\citenamefont {Harvey}\ \emph {et~al.}(2022)\citenamefont {Harvey}, \citenamefont {Wang}, \citenamefont {Fowlie}, \citenamefont {Osada}, \citenamefont {Lee}, \citenamefont {Lee}, \citenamefont {Li},\ and\ \citenamefont {Hwang}}]{harvey22}%
  \BibitemOpen
  \bibfield  {author} {\bibinfo {author} {\bibfnamefont {S.~P.}\ \bibnamefont {Harvey}}, \bibinfo {author} {\bibfnamefont {B.~Y.}\ \bibnamefont {Wang}}, \bibinfo {author} {\bibfnamefont {J.}~\bibnamefont {Fowlie}}, \bibinfo {author} {\bibfnamefont {M.}~\bibnamefont {Osada}}, \bibinfo {author} {\bibfnamefont {K.}~\bibnamefont {Lee}}, \bibinfo {author} {\bibfnamefont {Y.}~\bibnamefont {Lee}}, \bibinfo {author} {\bibfnamefont {D.}~\bibnamefont {Li}}, \ and\ \bibinfo {author} {\bibfnamefont {H.~Y.}\ \bibnamefont {Hwang}},\ }\href {https://arxiv.org/abs/2201.12971} {\enquote {\bibinfo {title} {Evidence for nodal superconductivity in infinite-layer nickelates},}\ } (\bibinfo {year} {2022})\BibitemShut {NoStop}%
\bibitem [{\citenamefont {Wang}\ \emph {et~al.}(2020)\citenamefont {Wang}, \citenamefont {Zhang}, \citenamefont {Yang},\ and\ \citenamefont {Zhang}}]{zfc20}%
  \BibitemOpen
  \bibfield  {author} {\bibinfo {author} {\bibfnamefont {Z.}~\bibnamefont {Wang}}, \bibinfo {author} {\bibfnamefont {G.-M.}\ \bibnamefont {Zhang}}, \bibinfo {author} {\bibfnamefont {Y.-F.}\ \bibnamefont {Yang}}, \ and\ \bibinfo {author} {\bibfnamefont {F.-C.}\ \bibnamefont {Zhang}},\ }\href {\doibase 10.1103/physrevb.102.220501} {\bibfield  {journal} {\bibinfo  {journal} {Physical Review B}\ }\textbf {\bibinfo {volume} {102}} (\bibinfo {year} {2020}),\ 10.1103/physrevb.102.220501}\BibitemShut {NoStop}%
\bibitem [{\citenamefont {Zaanen}\ \emph {et~al.}(1985)\citenamefont {Zaanen}, \citenamefont {Sawatzky},\ and\ \citenamefont {Allen}}]{zaanen85}%
  \BibitemOpen
  \bibfield  {author} {\bibinfo {author} {\bibfnamefont {J.}~\bibnamefont {Zaanen}}, \bibinfo {author} {\bibfnamefont {G.~A.}\ \bibnamefont {Sawatzky}}, \ and\ \bibinfo {author} {\bibfnamefont {J.~W.}\ \bibnamefont {Allen}},\ }\href {\doibase 10.1103/PhysRevLett.55.418} {\bibfield  {journal} {\bibinfo  {journal} {Phys. Rev. Lett.}\ }\textbf {\bibinfo {volume} {55}},\ \bibinfo {pages} {418} (\bibinfo {year} {1985})}\BibitemShut {NoStop}%
\bibitem [{\citenamefont {Jiang}\ \emph {et~al.}(2020)\citenamefont {Jiang}, \citenamefont {Berciu},\ and\ \citenamefont {Sawatzky}}]{jiangm20}%
  \BibitemOpen
  \bibfield  {author} {\bibinfo {author} {\bibfnamefont {M.}~\bibnamefont {Jiang}}, \bibinfo {author} {\bibfnamefont {M.}~\bibnamefont {Berciu}}, \ and\ \bibinfo {author} {\bibfnamefont {G.~A.}\ \bibnamefont {Sawatzky}},\ }\href {\doibase 10.1103/PhysRevLett.124.207004} {\bibfield  {journal} {\bibinfo  {journal} {Phys Rev Lett}\ }\textbf {\bibinfo {volume} {124}},\ \bibinfo {pages} {207004} (\bibinfo {year} {2020})}\BibitemShut {NoStop}%
\bibitem [{\citenamefont {Been}\ \emph {et~al.}(2021)\citenamefont {Been}, \citenamefont {Lee}, \citenamefont {Hwang}, \citenamefont {Cui}, \citenamefont {Zaanen}, \citenamefont {Devereaux}, \citenamefont {Moritz},\ and\ \citenamefont {Jia}}]{emily20}%
  \BibitemOpen
  \bibfield  {author} {\bibinfo {author} {\bibfnamefont {E.}~\bibnamefont {Been}}, \bibinfo {author} {\bibfnamefont {W.-S.}\ \bibnamefont {Lee}}, \bibinfo {author} {\bibfnamefont {H.~Y.}\ \bibnamefont {Hwang}}, \bibinfo {author} {\bibfnamefont {Y.}~\bibnamefont {Cui}}, \bibinfo {author} {\bibfnamefont {J.}~\bibnamefont {Zaanen}}, \bibinfo {author} {\bibfnamefont {T.}~\bibnamefont {Devereaux}}, \bibinfo {author} {\bibfnamefont {B.}~\bibnamefont {Moritz}}, \ and\ \bibinfo {author} {\bibfnamefont {C.}~\bibnamefont {Jia}},\ }\href {\doibase 10.1103/PhysRevX.11.011050} {\bibfield  {journal} {\bibinfo  {journal} {Phys. Rev. X}\ }\textbf {\bibinfo {volume} {11}},\ \bibinfo {pages} {011050} (\bibinfo {year} {2021})}\BibitemShut {NoStop}%
\bibitem [{\citenamefont {Hepting}\ \emph {et~al.}(2020)\citenamefont {Hepting}, \citenamefont {Li}, \citenamefont {Jia}, \citenamefont {Lu}, \citenamefont {Paris}, \citenamefont {Tseng}, \citenamefont {Feng}, \citenamefont {Osada}, \citenamefont {Been}, \citenamefont {Hikita}, \citenamefont {Chuang}, \citenamefont {Hussain}, \citenamefont {Zhou}, \citenamefont {Nag}, \citenamefont {Garcia-Fernandez}, \citenamefont {Rossi}, \citenamefont {Huang}, \citenamefont {Huang}, \citenamefont {Shen}, \citenamefont {Schmitt}, \citenamefont {Hwang}, \citenamefont {Moritz}, \citenamefont {Zaanen}, \citenamefont {Devereaux},\ and\ \citenamefont {Lee}}]{hepting20}%
  \BibitemOpen
  \bibfield  {author} {\bibinfo {author} {\bibfnamefont {M.}~\bibnamefont {Hepting}}, \bibinfo {author} {\bibfnamefont {D.}~\bibnamefont {Li}}, \bibinfo {author} {\bibfnamefont {C.~J.}\ \bibnamefont {Jia}}, \bibinfo {author} {\bibfnamefont {H.}~\bibnamefont {Lu}}, \bibinfo {author} {\bibfnamefont {E.}~\bibnamefont {Paris}}, \bibinfo {author} {\bibfnamefont {Y.}~\bibnamefont {Tseng}}, \bibinfo {author} {\bibfnamefont {X.}~\bibnamefont {Feng}}, \bibinfo {author} {\bibfnamefont {M.}~\bibnamefont {Osada}}, \bibinfo {author} {\bibfnamefont {E.}~\bibnamefont {Been}}, \bibinfo {author} {\bibfnamefont {Y.}~\bibnamefont {Hikita}}, \bibinfo {author} {\bibfnamefont {Y.~D.}\ \bibnamefont {Chuang}}, \bibinfo {author} {\bibfnamefont {Z.}~\bibnamefont {Hussain}}, \bibinfo {author} {\bibfnamefont {K.~J.}\ \bibnamefont {Zhou}}, \bibinfo {author} {\bibfnamefont {A.}~\bibnamefont {Nag}}, \bibinfo {author} {\bibfnamefont {M.}~\bibnamefont {Garcia-Fernandez}}, \bibinfo {author} {\bibfnamefont {M.}~\bibnamefont {Rossi}}, \bibinfo
  {author} {\bibfnamefont {H.~Y.}\ \bibnamefont {Huang}}, \bibinfo {author} {\bibfnamefont {D.~J.}\ \bibnamefont {Huang}}, \bibinfo {author} {\bibfnamefont {Z.~X.}\ \bibnamefont {Shen}}, \bibinfo {author} {\bibfnamefont {T.}~\bibnamefont {Schmitt}}, \bibinfo {author} {\bibfnamefont {H.~Y.}\ \bibnamefont {Hwang}}, \bibinfo {author} {\bibfnamefont {B.}~\bibnamefont {Moritz}}, \bibinfo {author} {\bibfnamefont {J.}~\bibnamefont {Zaanen}}, \bibinfo {author} {\bibfnamefont {T.~P.}\ \bibnamefont {Devereaux}}, \ and\ \bibinfo {author} {\bibfnamefont {W.~S.}\ \bibnamefont {Lee}},\ }\href {\doibase 10.1038/s41563-019-0585-z} {\bibfield  {journal} {\bibinfo  {journal} {Nat Mater}\ }\textbf {\bibinfo {volume} {19}},\ \bibinfo {pages} {381} (\bibinfo {year} {2020})}\BibitemShut {NoStop}%
\bibitem [{\citenamefont {Botana}\ and\ \citenamefont {Norman}(2020)}]{botana20}%
  \BibitemOpen
  \bibfield  {author} {\bibinfo {author} {\bibfnamefont {A.~S.}\ \bibnamefont {Botana}}\ and\ \bibinfo {author} {\bibfnamefont {M.~R.}\ \bibnamefont {Norman}},\ }\href {\doibase 10.1103/PhysRevX.10.011024} {\bibfield  {journal} {\bibinfo  {journal} {Physical Review X}\ }\textbf {\bibinfo {volume} {10}} (\bibinfo {year} {2020}),\ 10.1103/PhysRevX.10.011024}\BibitemShut {NoStop}%
\bibitem [{\citenamefont {Zhang}\ \emph {et~al.}(2020{\natexlab{a}})\citenamefont {Zhang}, \citenamefont {Yang},\ and\ \citenamefont {Zhang}}]{zfc201}%
  \BibitemOpen
  \bibfield  {author} {\bibinfo {author} {\bibfnamefont {G.-M.}\ \bibnamefont {Zhang}}, \bibinfo {author} {\bibfnamefont {Y.-f.}\ \bibnamefont {Yang}}, \ and\ \bibinfo {author} {\bibfnamefont {F.-C.}\ \bibnamefont {Zhang}},\ }\href {\doibase 10.1103/PhysRevB.101.020501} {\bibfield  {journal} {\bibinfo  {journal} {Physical Review B}\ }\textbf {\bibinfo {volume} {101}} (\bibinfo {year} {2020}{\natexlab{a}}),\ 10.1103/PhysRevB.101.020501}\BibitemShut {NoStop}%
\bibitem [{\citenamefont {Goodge}\ \emph {et~al.}(2021)\citenamefont {Goodge}, \citenamefont {Li}, \citenamefont {Lee}, \citenamefont {Osada}, \citenamefont {Wang}, \citenamefont {Sawatzky}, \citenamefont {Hwang},\ and\ \citenamefont {Kourkoutis}}]{goodge21}%
  \BibitemOpen
  \bibfield  {author} {\bibinfo {author} {\bibfnamefont {B.~H.}\ \bibnamefont {Goodge}}, \bibinfo {author} {\bibfnamefont {D.}~\bibnamefont {Li}}, \bibinfo {author} {\bibfnamefont {K.}~\bibnamefont {Lee}}, \bibinfo {author} {\bibfnamefont {M.}~\bibnamefont {Osada}}, \bibinfo {author} {\bibfnamefont {B.~Y.}\ \bibnamefont {Wang}}, \bibinfo {author} {\bibfnamefont {G.~A.}\ \bibnamefont {Sawatzky}}, \bibinfo {author} {\bibfnamefont {H.~Y.}\ \bibnamefont {Hwang}}, \ and\ \bibinfo {author} {\bibfnamefont {L.~F.}\ \bibnamefont {Kourkoutis}},\ }\href {\doibase 10.1073/pnas.2007683118} {\bibfield  {journal} {\bibinfo  {journal} {Proc Natl Acad Sci U S A}\ }\textbf {\bibinfo {volume} {118}} (\bibinfo {year} {2021}),\ 10.1073/pnas.2007683118}\BibitemShut {NoStop}%
\bibitem [{\citenamefont {Petocchi}\ \emph {et~al.}(2020)\citenamefont {Petocchi}, \citenamefont {Christiansson}, \citenamefont {Nilsson}, \citenamefont {Aryasetiawan},\ and\ \citenamefont {Werner}}]{petocchi20}%
  \BibitemOpen
  \bibfield  {author} {\bibinfo {author} {\bibfnamefont {F.}~\bibnamefont {Petocchi}}, \bibinfo {author} {\bibfnamefont {V.}~\bibnamefont {Christiansson}}, \bibinfo {author} {\bibfnamefont {F.}~\bibnamefont {Nilsson}}, \bibinfo {author} {\bibfnamefont {F.}~\bibnamefont {Aryasetiawan}}, \ and\ \bibinfo {author} {\bibfnamefont {P.}~\bibnamefont {Werner}},\ }\href {\doibase 10.1103/PhysRevX.10.041047} {\bibfield  {journal} {\bibinfo  {journal} {Physical Review X}\ }\textbf {\bibinfo {volume} {10}} (\bibinfo {year} {2020}),\ 10.1103/PhysRevX.10.041047}\BibitemShut {NoStop}%
\bibitem [{\citenamefont {Leonov}\ \emph {et~al.}(2020)\citenamefont {Leonov}, \citenamefont {Skornyakov},\ and\ \citenamefont {Savrasov}}]{leonov20}%
  \BibitemOpen
  \bibfield  {author} {\bibinfo {author} {\bibfnamefont {I.}~\bibnamefont {Leonov}}, \bibinfo {author} {\bibfnamefont {S.~L.}\ \bibnamefont {Skornyakov}}, \ and\ \bibinfo {author} {\bibfnamefont {S.~Y.}\ \bibnamefont {Savrasov}},\ }\href {\doibase 10.1103/physrevb.101.241108} {\bibfield  {journal} {\bibinfo  {journal} {Physical Review B}\ }\textbf {\bibinfo {volume} {101}} (\bibinfo {year} {2020}),\ 10.1103/physrevb.101.241108}\BibitemShut {NoStop}%
\bibitem [{\citenamefont {Karp}\ \emph {et~al.}(2020)\citenamefont {Karp}, \citenamefont {Botana}, \citenamefont {Norman}, \citenamefont {Park}, \citenamefont {Zingl},\ and\ \citenamefont {Millis}}]{karp20}%
  \BibitemOpen
  \bibfield  {author} {\bibinfo {author} {\bibfnamefont {J.}~\bibnamefont {Karp}}, \bibinfo {author} {\bibfnamefont {A.~S.}\ \bibnamefont {Botana}}, \bibinfo {author} {\bibfnamefont {M.~R.}\ \bibnamefont {Norman}}, \bibinfo {author} {\bibfnamefont {H.}~\bibnamefont {Park}}, \bibinfo {author} {\bibfnamefont {M.}~\bibnamefont {Zingl}}, \ and\ \bibinfo {author} {\bibfnamefont {A.}~\bibnamefont {Millis}},\ }\href {\doibase 10.1103/physrevx.10.021061} {\bibfield  {journal} {\bibinfo  {journal} {Physical Review X}\ }\textbf {\bibinfo {volume} {10}} (\bibinfo {year} {2020}),\ 10.1103/physrevx.10.021061}\BibitemShut {NoStop}%
\bibitem [{\citenamefont {Lechermann}(2020{\natexlab{a}})}]{lechermann20_prx}%
  \BibitemOpen
  \bibfield  {author} {\bibinfo {author} {\bibfnamefont {F.}~\bibnamefont {Lechermann}},\ }\href {\doibase 10.1103/PhysRevX.10.041002} {\bibfield  {journal} {\bibinfo  {journal} {Physical Review X}\ }\textbf {\bibinfo {volume} {10}} (\bibinfo {year} {2020}{\natexlab{a}}),\ 10.1103/PhysRevX.10.041002}\BibitemShut {NoStop}%
\bibitem [{\citenamefont {Lechermann}(2020{\natexlab{b}})}]{lechermann20}%
  \BibitemOpen
  \bibfield  {author} {\bibinfo {author} {\bibfnamefont {F.}~\bibnamefont {Lechermann}},\ }\href {\doibase 10.1103/PhysRevB.101.081110} {\bibfield  {journal} {\bibinfo  {journal} {Physical Review B}\ }\textbf {\bibinfo {volume} {101}} (\bibinfo {year} {2020}{\natexlab{b}}),\ 10.1103/PhysRevB.101.081110}\BibitemShut {NoStop}%
\bibitem [{\citenamefont {Lechermann}(2021)}]{lechermann21}%
  \BibitemOpen
  \bibfield  {author} {\bibinfo {author} {\bibfnamefont {F.}~\bibnamefont {Lechermann}},\ }\href {\doibase 10.1103/physrevmaterials.5.044803} {\bibfield  {journal} {\bibinfo  {journal} {Physical Review Materials}\ }\textbf {\bibinfo {volume} {5}} (\bibinfo {year} {2021}),\ 10.1103/physrevmaterials.5.044803}\BibitemShut {NoStop}%
\bibitem [{\citenamefont {Hu}\ and\ \citenamefont {Ding}(2012{\natexlab{a}})}]{hujp12}%
  \BibitemOpen
  \bibfield  {author} {\bibinfo {author} {\bibfnamefont {J.}~\bibnamefont {Hu}}\ and\ \bibinfo {author} {\bibfnamefont {H.}~\bibnamefont {Ding}},\ }\href {\doibase 10.1038/srep00381} {\bibfield  {journal} {\bibinfo  {journal} {Scientific Reports}\ }\textbf {\bibinfo {volume} {2}} (\bibinfo {year} {2012}{\natexlab{a}}),\ 10.1038/srep00381}\BibitemShut {NoStop}%
\bibitem [{\citenamefont {Bulut}\ \emph {et~al.}(1992)\citenamefont {Bulut}, \citenamefont {Scalapino},\ and\ \citenamefont {Scalettar}}]{bulut92}%
  \BibitemOpen
  \bibfield  {author} {\bibinfo {author} {\bibfnamefont {N.}~\bibnamefont {Bulut}}, \bibinfo {author} {\bibfnamefont {D.~J.}\ \bibnamefont {Scalapino}}, \ and\ \bibinfo {author} {\bibfnamefont {R.~T.}\ \bibnamefont {Scalettar}},\ }\href {\doibase 10.1103/physrevb.45.5577} {\bibfield  {journal} {\bibinfo  {journal} {Phys Rev B Condens Matter}\ }\textbf {\bibinfo {volume} {45}},\ \bibinfo {pages} {5577} (\bibinfo {year} {1992})}\BibitemShut {NoStop}%
\bibitem [{\citenamefont {Hetzel}\ \emph {et~al.}(1994)\citenamefont {Hetzel}, \citenamefont {von~der Linden},\ and\ \citenamefont {Hanke}}]{hetzel94}%
  \BibitemOpen
  \bibfield  {author} {\bibinfo {author} {\bibfnamefont {R.~E.}\ \bibnamefont {Hetzel}}, \bibinfo {author} {\bibfnamefont {W.}~\bibnamefont {von~der Linden}}, \ and\ \bibinfo {author} {\bibfnamefont {W.}~\bibnamefont {Hanke}},\ }\href {\doibase 10.1103/physrevb.50.4159} {\bibfield  {journal} {\bibinfo  {journal} {Phys Rev B Condens Matter}\ }\textbf {\bibinfo {volume} {50}},\ \bibinfo {pages} {4159} (\bibinfo {year} {1994})}\BibitemShut {NoStop}%
\bibitem [{\citenamefont {Scalettar}\ \emph {et~al.}(1994)\citenamefont {Scalettar}, \citenamefont {Cannon}, \citenamefont {Scalapino},\ and\ \citenamefont {Sugar}}]{scalettar94}%
  \BibitemOpen
  \bibfield  {author} {\bibinfo {author} {\bibfnamefont {R.~T.}\ \bibnamefont {Scalettar}}, \bibinfo {author} {\bibfnamefont {J.~W.}\ \bibnamefont {Cannon}}, \bibinfo {author} {\bibfnamefont {D.~J.}\ \bibnamefont {Scalapino}}, \ and\ \bibinfo {author} {\bibfnamefont {R.~L.}\ \bibnamefont {Sugar}},\ }\href {\doibase 10.1103/physrevb.50.13419} {\bibfield  {journal} {\bibinfo  {journal} {Phys Rev B Condens Matter}\ }\textbf {\bibinfo {volume} {50}},\ \bibinfo {pages} {13419} (\bibinfo {year} {1994})}\BibitemShut {NoStop}%
\bibitem [{\citenamefont {dos Santos}(1995)}]{dos95}%
  \BibitemOpen
  \bibfield  {author} {\bibinfo {author} {\bibfnamefont {R.~R.}\ \bibnamefont {dos Santos}},\ }\href {\doibase 10.1103/physrevb.51.15540} {\bibfield  {journal} {\bibinfo  {journal} {Phys Rev B Condens Matter}\ }\textbf {\bibinfo {volume} {51}},\ \bibinfo {pages} {15540} (\bibinfo {year} {1995})}\BibitemShut {NoStop}%
\bibitem [{\citenamefont {Bouadim}\ \emph {et~al.}(2008)\citenamefont {Bouadim}, \citenamefont {Batrouni}, \citenamefont {Hébert},\ and\ \citenamefont {Scalettar}}]{bouadim08}%
  \BibitemOpen
  \bibfield  {author} {\bibinfo {author} {\bibfnamefont {K.}~\bibnamefont {Bouadim}}, \bibinfo {author} {\bibfnamefont {G.~G.}\ \bibnamefont {Batrouni}}, \bibinfo {author} {\bibfnamefont {F.}~\bibnamefont {Hébert}}, \ and\ \bibinfo {author} {\bibfnamefont {R.~T.}\ \bibnamefont {Scalettar}},\ }\href {\doibase 10.1103/PhysRevB.77.144527} {\bibfield  {journal} {\bibinfo  {journal} {Physical Review B}\ }\textbf {\bibinfo {volume} {77}} (\bibinfo {year} {2008}),\ 10.1103/PhysRevB.77.144527}\BibitemShut {NoStop}%
\bibitem [{\citenamefont {Kancharla}\ and\ \citenamefont {Okamoto}(2007)}]{kancharla07}%
  \BibitemOpen
  \bibfield  {author} {\bibinfo {author} {\bibfnamefont {S.~S.}\ \bibnamefont {Kancharla}}\ and\ \bibinfo {author} {\bibfnamefont {S.}~\bibnamefont {Okamoto}},\ }\href {\doibase 10.1103/PhysRevB.75.193103} {\bibfield  {journal} {\bibinfo  {journal} {Physical Review B}\ }\textbf {\bibinfo {volume} {75}} (\bibinfo {year} {2007}),\ 10.1103/PhysRevB.75.193103}\BibitemShut {NoStop}%
\bibitem [{\citenamefont {Maier}\ and\ \citenamefont {Scalapino}(2011)}]{maier11}%
  \BibitemOpen
  \bibfield  {author} {\bibinfo {author} {\bibfnamefont {T.~A.}\ \bibnamefont {Maier}}\ and\ \bibinfo {author} {\bibfnamefont {D.~J.}\ \bibnamefont {Scalapino}},\ }\href {\doibase 10.1103/physrevb.84.180513} {\bibfield  {journal} {\bibinfo  {journal} {Physical Review B}\ }\textbf {\bibinfo {volume} {84}} (\bibinfo {year} {2011}),\ 10.1103/physrevb.84.180513}\BibitemShut {NoStop}%
\bibitem [{\citenamefont {Mishra}\ \emph {et~al.}(2016)\citenamefont {Mishra}, \citenamefont {Scalapino},\ and\ \citenamefont {Maier}}]{mishra16}%
  \BibitemOpen
  \bibfield  {author} {\bibinfo {author} {\bibfnamefont {V.}~\bibnamefont {Mishra}}, \bibinfo {author} {\bibfnamefont {D.~J.}\ \bibnamefont {Scalapino}}, \ and\ \bibinfo {author} {\bibfnamefont {T.~A.}\ \bibnamefont {Maier}},\ }\href {\doibase 10.1038/srep32078} {\bibfield  {journal} {\bibinfo  {journal} {Sci Rep}\ }\textbf {\bibinfo {volume} {6}},\ \bibinfo {pages} {32078} (\bibinfo {year} {2016})}\BibitemShut {NoStop}%
\bibitem [{\citenamefont {Nakata}\ \emph {et~al.}(2017)\citenamefont {Nakata}, \citenamefont {Ogura}, \citenamefont {Usui},\ and\ \citenamefont {Kuroki}}]{nakata17}%
  \BibitemOpen
  \bibfield  {author} {\bibinfo {author} {\bibfnamefont {M.}~\bibnamefont {Nakata}}, \bibinfo {author} {\bibfnamefont {D.}~\bibnamefont {Ogura}}, \bibinfo {author} {\bibfnamefont {H.}~\bibnamefont {Usui}}, \ and\ \bibinfo {author} {\bibfnamefont {K.}~\bibnamefont {Kuroki}},\ }\href {\doibase 10.1103/PhysRevB.95.214509} {\bibfield  {journal} {\bibinfo  {journal} {Physical Review B}\ }\textbf {\bibinfo {volume} {95}} (\bibinfo {year} {2017}),\ 10.1103/PhysRevB.95.214509}\BibitemShut {NoStop}%
\bibitem [{\citenamefont {Maier}\ \emph {et~al.}(2019)\citenamefont {Maier}, \citenamefont {Mishra}, \citenamefont {Balduzzi},\ and\ \citenamefont {Scalapino}}]{maier19}%
  \BibitemOpen
  \bibfield  {author} {\bibinfo {author} {\bibfnamefont {T.~A.}\ \bibnamefont {Maier}}, \bibinfo {author} {\bibfnamefont {V.}~\bibnamefont {Mishra}}, \bibinfo {author} {\bibfnamefont {G.}~\bibnamefont {Balduzzi}}, \ and\ \bibinfo {author} {\bibfnamefont {D.~J.}\ \bibnamefont {Scalapino}},\ }\href {\doibase 10.1103/PhysRevB.99.140504} {\bibfield  {journal} {\bibinfo  {journal} {Physical Review B}\ }\textbf {\bibinfo {volume} {99}} (\bibinfo {year} {2019}),\ 10.1103/PhysRevB.99.140504}\BibitemShut {NoStop}%
\bibitem [{\citenamefont {Karakuzu}\ \emph {et~al.}(2021)\citenamefont {Karakuzu}, \citenamefont {Johnston},\ and\ \citenamefont {Maier}}]{karakuzu21}%
  \BibitemOpen
  \bibfield  {author} {\bibinfo {author} {\bibfnamefont {S.}~\bibnamefont {Karakuzu}}, \bibinfo {author} {\bibfnamefont {S.}~\bibnamefont {Johnston}}, \ and\ \bibinfo {author} {\bibfnamefont {T.~A.}\ \bibnamefont {Maier}},\ }\href {\doibase 10.1103/PhysRevB.104.245109} {\bibfield  {journal} {\bibinfo  {journal} {Physical Review B}\ }\textbf {\bibinfo {volume} {104}} (\bibinfo {year} {2021}),\ 10.1103/PhysRevB.104.245109}\BibitemShut {NoStop}%
\bibitem [{\citenamefont {Nica}\ and\ \citenamefont {Erten}(2020)}]{nica20}%
  \BibitemOpen
  \bibfield  {author} {\bibinfo {author} {\bibfnamefont {E.~M.}\ \bibnamefont {Nica}}\ and\ \bibinfo {author} {\bibfnamefont {O.}~\bibnamefont {Erten}},\ }\href {\doibase 10.1103/PhysRevB.102.214509} {\bibfield  {journal} {\bibinfo  {journal} {Physical Review B}\ }\textbf {\bibinfo {volume} {102}} (\bibinfo {year} {2020}),\ 10.1103/PhysRevB.102.214509}\BibitemShut {NoStop}%
\bibitem [{\citenamefont {Wu}\ \emph {et~al.}(2020{\natexlab{a}})\citenamefont {Wu}, \citenamefont {Jiang}, \citenamefont {Sante}, \citenamefont {Hanke}, \citenamefont {Schnyder}, \citenamefont {Hu},\ and\ \citenamefont {Thomale}}]{wuxx20}%
  \BibitemOpen
  \bibfield  {author} {\bibinfo {author} {\bibfnamefont {X.}~\bibnamefont {Wu}}, \bibinfo {author} {\bibfnamefont {K.}~\bibnamefont {Jiang}}, \bibinfo {author} {\bibfnamefont {D.~D.}\ \bibnamefont {Sante}}, \bibinfo {author} {\bibfnamefont {W.}~\bibnamefont {Hanke}}, \bibinfo {author} {\bibfnamefont {A.~P.}\ \bibnamefont {Schnyder}}, \bibinfo {author} {\bibfnamefont {J.}~\bibnamefont {Hu}}, \ and\ \bibinfo {author} {\bibfnamefont {R.}~\bibnamefont {Thomale}},\ }\href@noop {} {\enquote {\bibinfo {title} {Surface $s$-wave superconductivity for oxide-terminated infinite-layer nickelates},}\ } (\bibinfo {year} {2020}{\natexlab{a}}),\ \Eprint {http://arxiv.org/abs/2008.06009} {arXiv:2008.06009 [cond-mat.supr-con]} \BibitemShut {NoStop}%
\bibitem [{\citenamefont {Maier}(2005)}]{maier05}%
  \BibitemOpen
  \bibfield  {author} {\bibinfo {author} {\bibfnamefont {T.~A.}\ \bibnamefont {Maier}},\ }\href@noop {} {\bibfield  {journal} {\bibinfo  {journal} {Reviews of Modern Physics}\ }\textbf {\bibinfo {volume} {77}} (\bibinfo {year} {2005})}\BibitemShut {NoStop}%
\bibitem [{\citenamefont {Maier}\ \emph {et~al.}(2005)\citenamefont {Maier}, \citenamefont {Jarrell}, \citenamefont {Schulthess}, \citenamefont {Kent},\ and\ \citenamefont {White}}]{maier05_prl}%
  \BibitemOpen
  \bibfield  {author} {\bibinfo {author} {\bibfnamefont {T.~A.}\ \bibnamefont {Maier}}, \bibinfo {author} {\bibfnamefont {M.}~\bibnamefont {Jarrell}}, \bibinfo {author} {\bibfnamefont {T.~C.}\ \bibnamefont {Schulthess}}, \bibinfo {author} {\bibfnamefont {P.~R.~C.}\ \bibnamefont {Kent}}, \ and\ \bibinfo {author} {\bibfnamefont {J.~B.}\ \bibnamefont {White}},\ }\href {\doibase 10.1103/PhysRevLett.95.237001} {\bibfield  {journal} {\bibinfo  {journal} {Phys. Rev. Lett.}\ }\textbf {\bibinfo {volume} {95}},\ \bibinfo {pages} {237001} (\bibinfo {year} {2005})}\BibitemShut {NoStop}%
\bibitem [{\citenamefont {Fernandes}\ and\ \citenamefont {Chubukov}(2017)}]{fernandes17}%
  \BibitemOpen
  \bibfield  {author} {\bibinfo {author} {\bibfnamefont {R.~M.}\ \bibnamefont {Fernandes}}\ and\ \bibinfo {author} {\bibfnamefont {A.~V.}\ \bibnamefont {Chubukov}},\ }\href {\doibase 10.1088/1361-6633/80/1/014503} {\bibfield  {journal} {\bibinfo  {journal} {Rep Prog Phys}\ }\textbf {\bibinfo {volume} {80}},\ \bibinfo {pages} {014503} (\bibinfo {year} {2017})}\BibitemShut {NoStop}%
\bibitem [{\citenamefont {Georges}(1996)}]{georges96}%
  \BibitemOpen
  \bibfield  {author} {\bibinfo {author} {\bibfnamefont {A.}~\bibnamefont {Georges}},\ }\href@noop {} {\bibfield  {journal} {\bibinfo  {journal} {Reviews of Modern Physics}\ }\textbf {\bibinfo {volume} {68}} (\bibinfo {year} {1996})}\BibitemShut {NoStop}%
\bibitem [{\citenamefont {Kotliar}\ \emph {et~al.}(2001)\citenamefont {Kotliar}, \citenamefont {Savrasov}, \citenamefont {Pálsson},\ and\ \citenamefont {Biroli}}]{kotliar01}%
  \BibitemOpen
  \bibfield  {author} {\bibinfo {author} {\bibfnamefont {G.}~\bibnamefont {Kotliar}}, \bibinfo {author} {\bibfnamefont {S.~Y.}\ \bibnamefont {Savrasov}}, \bibinfo {author} {\bibfnamefont {G.}~\bibnamefont {Pálsson}}, \ and\ \bibinfo {author} {\bibfnamefont {G.}~\bibnamefont {Biroli}},\ }\href {\doibase 10.1103/physrevlett.87.186401} {\bibfield  {journal} {\bibinfo  {journal} {Physical Review Letters}\ }\textbf {\bibinfo {volume} {87}} (\bibinfo {year} {2001}),\ 10.1103/physrevlett.87.186401}\BibitemShut {NoStop}%
\bibitem [{\citenamefont {Hirsch}\ and\ \citenamefont {Fye}(1986)}]{hirschfye86}%
  \BibitemOpen
  \bibfield  {author} {\bibinfo {author} {\bibfnamefont {J.~E.}\ \bibnamefont {Hirsch}}\ and\ \bibinfo {author} {\bibfnamefont {R.~M.}\ \bibnamefont {Fye}},\ }\href {\doibase 10.1103/PhysRevLett.56.2521} {\bibfield  {journal} {\bibinfo  {journal} {Phys Rev Lett}\ }\textbf {\bibinfo {volume} {56}},\ \bibinfo {pages} {2521} (\bibinfo {year} {1986})}\BibitemShut {NoStop}%
\bibitem [{\citenamefont {Lin}\ \emph {et~al.}(2012)\citenamefont {Lin}, \citenamefont {Gull},\ and\ \citenamefont {Millis}}]{lin12}%
  \BibitemOpen
  \bibfield  {author} {\bibinfo {author} {\bibfnamefont {N.}~\bibnamefont {Lin}}, \bibinfo {author} {\bibfnamefont {E.}~\bibnamefont {Gull}}, \ and\ \bibinfo {author} {\bibfnamefont {A.~J.}\ \bibnamefont {Millis}},\ }\href {\doibase 10.1103/PhysRevLett.109.106401} {\bibfield  {journal} {\bibinfo  {journal} {Phys Rev Lett}\ }\textbf {\bibinfo {volume} {109}},\ \bibinfo {pages} {106401} (\bibinfo {year} {2012})}\BibitemShut {NoStop}%
\bibitem [{\citenamefont {Black-Schaffer}\ \emph {et~al.}(2014)\citenamefont {Black-Schaffer}, \citenamefont {Wu},\ and\ \citenamefont {Le~Hur}}]{ww14}%
  \BibitemOpen
  \bibfield  {author} {\bibinfo {author} {\bibfnamefont {A.~M.}\ \bibnamefont {Black-Schaffer}}, \bibinfo {author} {\bibfnamefont {W.}~\bibnamefont {Wu}}, \ and\ \bibinfo {author} {\bibfnamefont {K.}~\bibnamefont {Le~Hur}},\ }\href {\doibase 10.1103/PhysRevB.90.054521} {\bibfield  {journal} {\bibinfo  {journal} {Physical Review B}\ }\textbf {\bibinfo {volume} {90}} (\bibinfo {year} {2014}),\ 10.1103/PhysRevB.90.054521}\BibitemShut {NoStop}%
\bibitem [{\citenamefont {Wu}\ and\ \citenamefont {Tremblay}(2015)}]{wuwei15}%
  \BibitemOpen
  \bibfield  {author} {\bibinfo {author} {\bibfnamefont {W.}~\bibnamefont {Wu}}\ and\ \bibinfo {author} {\bibfnamefont {A.-M.-S.}\ \bibnamefont {Tremblay}},\ }\href {\doibase 10.1103/physrevx.5.011019} {\bibfield  {journal} {\bibinfo  {journal} {Physical Review X}\ }\textbf {\bibinfo {volume} {5}} (\bibinfo {year} {2015}),\ 10.1103/physrevx.5.011019}\BibitemShut {NoStop}%
\bibitem [{\citenamefont {Jiang}(2022)}]{jiang20}%
  \BibitemOpen
  \bibfield  {author} {\bibinfo {author} {\bibfnamefont {M.}~\bibnamefont {Jiang}},\ }\href@noop {} {\enquote {\bibinfo {title} {Characterizing the superconducting instability in a two-orbital $d$-$s$ model: insights to infinite-layer nickelate superconductors},}\ } (\bibinfo {year} {2022}),\ \Eprint {http://arxiv.org/abs/2201.12967} {arXiv:2201.12967 [cond-mat.supr-con]} \BibitemShut {NoStop}%
\bibitem [{\citenamefont {Zhang}\ and\ \citenamefont {Rice}(1988)}]{zrs}%
  \BibitemOpen
  \bibfield  {author} {\bibinfo {author} {\bibfnamefont {F.~C.}\ \bibnamefont {Zhang}}\ and\ \bibinfo {author} {\bibfnamefont {T.~M.}\ \bibnamefont {Rice}},\ }\href {\doibase 10.1103/PhysRevB.37.3759} {\bibfield  {journal} {\bibinfo  {journal} {Phys. Rev. B}\ }\textbf {\bibinfo {volume} {37}},\ \bibinfo {pages} {3759} (\bibinfo {year} {1988})}\BibitemShut {NoStop}%
\bibitem [{\citenamefont {Fratino}\ \emph {et~al.}(2016)\citenamefont {Fratino}, \citenamefont {Sémon}, \citenamefont {Sordi},\ and\ \citenamefont {Tremblay}}]{fratino16}%
  \BibitemOpen
  \bibfield  {author} {\bibinfo {author} {\bibfnamefont {L.}~\bibnamefont {Fratino}}, \bibinfo {author} {\bibfnamefont {P.}~\bibnamefont {Sémon}}, \bibinfo {author} {\bibfnamefont {G.}~\bibnamefont {Sordi}}, \ and\ \bibinfo {author} {\bibfnamefont {A.-M.~S.}\ \bibnamefont {Tremblay}},\ }\href {\doibase 10.1038/srep22715} {\bibfield  {journal} {\bibinfo  {journal} {Scientific Reports}\ }\textbf {\bibinfo {volume} {6}},\ \bibinfo {pages} {22715} (\bibinfo {year} {2016})}\BibitemShut {NoStop}%
\bibitem [{\citenamefont {Gunnarsson}\ \emph {et~al.}(2015)\citenamefont {Gunnarsson}, \citenamefont {Schafer}, \citenamefont {LeBlanc}, \citenamefont {Gull}, \citenamefont {Merino}, \citenamefont {Sangiovanni}, \citenamefont {Rohringer},\ and\ \citenamefont {Toschi}}]{Gunnarsson15}%
  \BibitemOpen
  \bibfield  {author} {\bibinfo {author} {\bibfnamefont {O.}~\bibnamefont {Gunnarsson}}, \bibinfo {author} {\bibfnamefont {T.}~\bibnamefont {Schafer}}, \bibinfo {author} {\bibfnamefont {J.~P.}\ \bibnamefont {LeBlanc}}, \bibinfo {author} {\bibfnamefont {E.}~\bibnamefont {Gull}}, \bibinfo {author} {\bibfnamefont {J.}~\bibnamefont {Merino}}, \bibinfo {author} {\bibfnamefont {G.}~\bibnamefont {Sangiovanni}}, \bibinfo {author} {\bibfnamefont {G.}~\bibnamefont {Rohringer}}, \ and\ \bibinfo {author} {\bibfnamefont {A.}~\bibnamefont {Toschi}},\ }\href {\doibase 10.1103/PhysRevLett.114.236402} {\bibfield  {journal} {\bibinfo  {journal} {Phys Rev Lett}\ }\textbf {\bibinfo {volume} {114}},\ \bibinfo {pages} {236402} (\bibinfo {year} {2015})}\BibitemShut {NoStop}%
\bibitem [{\citenamefont {Hu}\ and\ \citenamefont {Ding}(2012{\natexlab{b}})}]{hjp_SR2012}%
  \BibitemOpen
  \bibfield  {author} {\bibinfo {author} {\bibfnamefont {J.}~\bibnamefont {Hu}}\ and\ \bibinfo {author} {\bibfnamefont {H.}~\bibnamefont {Ding}},\ }\href {\doibase 10.1038/srep00381} {\bibfield  {journal} {\bibinfo  {journal} {Scientific Reports}\ }\textbf {\bibinfo {volume} {2}} (\bibinfo {year} {2012}{\natexlab{b}}),\ 10.1038/srep00381}\BibitemShut {NoStop}%
\bibitem [{\citenamefont {Wu}\ \emph {et~al.}(2018)\citenamefont {Wu}, \citenamefont {Scheurer}, \citenamefont {Chatterjee}, \citenamefont {Sachdev}, \citenamefont {Georges},\ and\ \citenamefont {Ferrero}}]{wuwei18}%
  \BibitemOpen
  \bibfield  {author} {\bibinfo {author} {\bibfnamefont {W.}~\bibnamefont {Wu}}, \bibinfo {author} {\bibfnamefont {M.~S.}\ \bibnamefont {Scheurer}}, \bibinfo {author} {\bibfnamefont {S.}~\bibnamefont {Chatterjee}}, \bibinfo {author} {\bibfnamefont {S.}~\bibnamefont {Sachdev}}, \bibinfo {author} {\bibfnamefont {A.}~\bibnamefont {Georges}}, \ and\ \bibinfo {author} {\bibfnamefont {M.}~\bibnamefont {Ferrero}},\ }\href {\doibase 10.1103/PhysRevX.8.021048} {\bibfield  {journal} {\bibinfo  {journal} {Physical Review X}\ }\textbf {\bibinfo {volume} {8}} (\bibinfo {year} {2018}),\ 10.1103/PhysRevX.8.021048}\BibitemShut {NoStop}%
\bibitem [{\citenamefont {Schafer}\ and\ \citenamefont {Toschi}(2021)}]{schafer21}%
  \BibitemOpen
  \bibfield  {author} {\bibinfo {author} {\bibfnamefont {T.}~\bibnamefont {Schafer}}\ and\ \bibinfo {author} {\bibfnamefont {A.}~\bibnamefont {Toschi}},\ }\href {\doibase 10.1088/1361-648X/abeb44} {\bibfield  {journal} {\bibinfo  {journal} {J Phys Condens Matter}\ }\textbf {\bibinfo {volume} {33}} (\bibinfo {year} {2021}),\ 10.1088/1361-648X/abeb44}\BibitemShut {NoStop}%
\bibitem [{\citenamefont {Dong}\ \emph {et~al.}(2022)\citenamefont {Dong}, \citenamefont {Re}, \citenamefont {Toschi},\ and\ \citenamefont {Gull}}]{Xinyang22}%
  \BibitemOpen
  \bibfield  {author} {\bibinfo {author} {\bibfnamefont {X.}~\bibnamefont {Dong}}, \bibinfo {author} {\bibfnamefont {L.~D.}\ \bibnamefont {Re}}, \bibinfo {author} {\bibfnamefont {A.}~\bibnamefont {Toschi}}, \ and\ \bibinfo {author} {\bibfnamefont {E.}~\bibnamefont {Gull}},\ }\href {\doibase 10.1073/pnas.2205048119} {\bibfield  {journal} {\bibinfo  {journal} {Proceedings of the National Academy of Sciences}\ }\textbf {\bibinfo {volume} {119}},\ \bibinfo {pages} {e2205048119} (\bibinfo {year} {2022})},\ \Eprint {http://arxiv.org/abs/https://www.pnas.org/doi/pdf/10.1073/pnas.2205048119} {https://www.pnas.org/doi/pdf/10.1073/pnas.2205048119} \BibitemShut {NoStop}%
\bibitem [{\citenamefont {Wu}\ \emph {et~al.}(2022)\citenamefont {Wu}, \citenamefont {Wang},\ and\ \citenamefont {Tremblay}}]{ww22}%
  \BibitemOpen
  \bibfield  {author} {\bibinfo {author} {\bibfnamefont {W.}~\bibnamefont {Wu}}, \bibinfo {author} {\bibfnamefont {X.}~\bibnamefont {Wang}}, \ and\ \bibinfo {author} {\bibfnamefont {A.~M.}\ \bibnamefont {Tremblay}},\ }\href {\doibase 10.1073/pnas.2115819119} {\bibfield  {journal} {\bibinfo  {journal} {Proc Natl Acad Sci U S A}\ }\textbf {\bibinfo {volume} {119}},\ \bibinfo {pages} {e2115819119} (\bibinfo {year} {2022})}\BibitemShut {NoStop}%
\bibitem [{\citenamefont {Wu}\ \emph {et~al.}(2020{\natexlab{b}})\citenamefont {Wu}, \citenamefont {Di~Sante}, \citenamefont {Schwemmer}, \citenamefont {Hanke}, \citenamefont {Hwang}, \citenamefont {Raghu},\ and\ \citenamefont {Thomale}}]{wuxx201}%
  \BibitemOpen
  \bibfield  {author} {\bibinfo {author} {\bibfnamefont {X.}~\bibnamefont {Wu}}, \bibinfo {author} {\bibfnamefont {D.}~\bibnamefont {Di~Sante}}, \bibinfo {author} {\bibfnamefont {T.}~\bibnamefont {Schwemmer}}, \bibinfo {author} {\bibfnamefont {W.}~\bibnamefont {Hanke}}, \bibinfo {author} {\bibfnamefont {H.~Y.}\ \bibnamefont {Hwang}}, \bibinfo {author} {\bibfnamefont {S.}~\bibnamefont {Raghu}}, \ and\ \bibinfo {author} {\bibfnamefont {R.}~\bibnamefont {Thomale}},\ }\href {\doibase 10.1103/PhysRevB.101.060504} {\bibfield  {journal} {\bibinfo  {journal} {Phys. Rev. B}\ }\textbf {\bibinfo {volume} {101}},\ \bibinfo {pages} {060504} (\bibinfo {year} {2020}{\natexlab{b}})}\BibitemShut {NoStop}%
\bibitem [{\citenamefont {Lu}\ \emph {et~al.}(2021)\citenamefont {Lu}, \citenamefont {Rossi}, \citenamefont {Nag}, \citenamefont {Osada}, \citenamefont {Li}, \citenamefont {Lee}, \citenamefont {Wang}, \citenamefont {Garcia-Fernandez}, \citenamefont {Agrestini}, \citenamefont {Shen}, \citenamefont {Been}, \citenamefont {Moritz}, \citenamefont {Devereaux}, \citenamefont {Zaanen}, \citenamefont {Hwang}, \citenamefont {Zhou},\ and\ \citenamefont {Lee}}]{luh21}%
  \BibitemOpen
  \bibfield  {author} {\bibinfo {author} {\bibfnamefont {H.}~\bibnamefont {Lu}}, \bibinfo {author} {\bibfnamefont {M.}~\bibnamefont {Rossi}}, \bibinfo {author} {\bibfnamefont {A.}~\bibnamefont {Nag}}, \bibinfo {author} {\bibfnamefont {M.}~\bibnamefont {Osada}}, \bibinfo {author} {\bibfnamefont {D.~F.}\ \bibnamefont {Li}}, \bibinfo {author} {\bibfnamefont {K.}~\bibnamefont {Lee}}, \bibinfo {author} {\bibfnamefont {B.~Y.}\ \bibnamefont {Wang}}, \bibinfo {author} {\bibfnamefont {M.}~\bibnamefont {Garcia-Fernandez}}, \bibinfo {author} {\bibfnamefont {S.}~\bibnamefont {Agrestini}}, \bibinfo {author} {\bibfnamefont {Z.~X.}\ \bibnamefont {Shen}}, \bibinfo {author} {\bibfnamefont {E.~M.}\ \bibnamefont {Been}}, \bibinfo {author} {\bibfnamefont {B.}~\bibnamefont {Moritz}}, \bibinfo {author} {\bibfnamefont {T.~P.}\ \bibnamefont {Devereaux}}, \bibinfo {author} {\bibfnamefont {J.}~\bibnamefont {Zaanen}}, \bibinfo {author} {\bibfnamefont {H.~Y.}\ \bibnamefont {Hwang}}, \bibinfo {author} {\bibfnamefont {K.~J.}\ \bibnamefont
  {Zhou}}, \ and\ \bibinfo {author} {\bibfnamefont {W.~S.}\ \bibnamefont {Lee}},\ }\href {\doibase 10.1126/science.abd7726} {\bibfield  {journal} {\bibinfo  {journal} {Science}\ }\textbf {\bibinfo {volume} {373}},\ \bibinfo {pages} {213} (\bibinfo {year} {2021})}\BibitemShut {NoStop}%
\bibitem [{\citenamefont {Zhang}\ \emph {et~al.}(2020{\natexlab{b}})\citenamefont {Zhang}, \citenamefont {Jin}, \citenamefont {Wang}, \citenamefont {Xi}, \citenamefont {Shi}, \citenamefont {Ye},\ and\ \citenamefont {Mei}}]{zhangh20}%
  \BibitemOpen
  \bibfield  {author} {\bibinfo {author} {\bibfnamefont {H.}~\bibnamefont {Zhang}}, \bibinfo {author} {\bibfnamefont {L.}~\bibnamefont {Jin}}, \bibinfo {author} {\bibfnamefont {S.}~\bibnamefont {Wang}}, \bibinfo {author} {\bibfnamefont {B.}~\bibnamefont {Xi}}, \bibinfo {author} {\bibfnamefont {X.}~\bibnamefont {Shi}}, \bibinfo {author} {\bibfnamefont {F.}~\bibnamefont {Ye}}, \ and\ \bibinfo {author} {\bibfnamefont {J.-W.}\ \bibnamefont {Mei}},\ }\href {\doibase 10.1103/physrevresearch.2.013214} {\bibfield  {journal} {\bibinfo  {journal} {Physical Review Research}\ }\textbf {\bibinfo {volume} {2}} (\bibinfo {year} {2020}{\natexlab{b}}),\ 10.1103/physrevresearch.2.013214}\BibitemShut {NoStop}%
\bibitem [{\citenamefont {Choi}\ \emph {et~al.}(2020)\citenamefont {Choi}, \citenamefont {Lee},\ and\ \citenamefont {Pickett}}]{choi20}%
  \BibitemOpen
  \bibfield  {author} {\bibinfo {author} {\bibfnamefont {M.-Y.}\ \bibnamefont {Choi}}, \bibinfo {author} {\bibfnamefont {K.-W.}\ \bibnamefont {Lee}}, \ and\ \bibinfo {author} {\bibfnamefont {W.~E.}\ \bibnamefont {Pickett}},\ }\href {\doibase 10.1103/PhysRevB.101.020503} {\bibfield  {journal} {\bibinfo  {journal} {Physical Review B}\ }\textbf {\bibinfo {volume} {101}} (\bibinfo {year} {2020}),\ 10.1103/PhysRevB.101.020503}\BibitemShut {NoStop}%
\bibitem [{\citenamefont {Huang}\ \emph {et~al.}(2022)\citenamefont {Huang}, \citenamefont {Yu}, \citenamefont {Xu}, \citenamefont {Zhu}, \citenamefont {Oh}, \citenamefont {Jiang}, \citenamefont {Wang}, \citenamefont {Wu}, \citenamefont {Chen}, \citenamefont {Denlinger}, \citenamefont {Mo}, \citenamefont {Hashimoto}, \citenamefont {Michiardi}, \citenamefont {Pedersen}, \citenamefont {Gorovikov}, \citenamefont {Zhdanovich}, \citenamefont {Damascelli}, \citenamefont {Gu}, \citenamefont {Dai}, \citenamefont {Chu}, \citenamefont {Lu}, \citenamefont {Si}, \citenamefont {Birgeneau},\ and\ \citenamefont {Yi}}]{huangjw22}%
  \BibitemOpen
  \bibfield  {author} {\bibinfo {author} {\bibfnamefont {J.}~\bibnamefont {Huang}}, \bibinfo {author} {\bibfnamefont {R.}~\bibnamefont {Yu}}, \bibinfo {author} {\bibfnamefont {Z.}~\bibnamefont {Xu}}, \bibinfo {author} {\bibfnamefont {J.-X.}\ \bibnamefont {Zhu}}, \bibinfo {author} {\bibfnamefont {J.~S.}\ \bibnamefont {Oh}}, \bibinfo {author} {\bibfnamefont {Q.}~\bibnamefont {Jiang}}, \bibinfo {author} {\bibfnamefont {M.}~\bibnamefont {Wang}}, \bibinfo {author} {\bibfnamefont {H.}~\bibnamefont {Wu}}, \bibinfo {author} {\bibfnamefont {T.}~\bibnamefont {Chen}}, \bibinfo {author} {\bibfnamefont {J.~D.}\ \bibnamefont {Denlinger}}, \bibinfo {author} {\bibfnamefont {S.-K.}\ \bibnamefont {Mo}}, \bibinfo {author} {\bibfnamefont {M.}~\bibnamefont {Hashimoto}}, \bibinfo {author} {\bibfnamefont {M.}~\bibnamefont {Michiardi}}, \bibinfo {author} {\bibfnamefont {T.~M.}\ \bibnamefont {Pedersen}}, \bibinfo {author} {\bibfnamefont {S.}~\bibnamefont {Gorovikov}}, \bibinfo {author} {\bibfnamefont {S.}~\bibnamefont {Zhdanovich}},
  \bibinfo {author} {\bibfnamefont {A.}~\bibnamefont {Damascelli}}, \bibinfo {author} {\bibfnamefont {G.}~\bibnamefont {Gu}}, \bibinfo {author} {\bibfnamefont {P.}~\bibnamefont {Dai}}, \bibinfo {author} {\bibfnamefont {J.-H.}\ \bibnamefont {Chu}}, \bibinfo {author} {\bibfnamefont {D.}~\bibnamefont {Lu}}, \bibinfo {author} {\bibfnamefont {Q.}~\bibnamefont {Si}}, \bibinfo {author} {\bibfnamefont {R.~J.}\ \bibnamefont {Birgeneau}}, \ and\ \bibinfo {author} {\bibfnamefont {M.}~\bibnamefont {Yi}},\ }\href {\doibase 10.1038/s42005-022-00805-6} {\bibfield  {journal} {\bibinfo  {journal} {Communications Physics}\ }\textbf {\bibinfo {volume} {5}} (\bibinfo {year} {2022}),\ 10.1038/s42005-022-00805-6}\BibitemShut {NoStop}%
\end{thebibliography}%

\end{document}